\documentclass[prl,aps,10pt,longbibliography,superscriptaddress,twocolumn]{revtex4-2}%
\usepackage{amsmath}
\usepackage{amssymb}
\usepackage{amsthm}
\usepackage{amsfonts}
\usepackage{enumerate}
\usepackage{latexsym}
\usepackage{bm}
\usepackage{graphicx}
\usepackage{subfigure}
\usepackage{color}
\usepackage[dvipsnames]{xcolor}
\usepackage[normalem]{ulem}
\usepackage[colorlinks]{hyperref}
\hypersetup{
  colorlinks,
  citecolor=Violet,
  linkcolor=Red,
  urlcolor=Blue}
\usepackage{ulem}
\usepackage{cancel}

\newcommand{\beq}{\begin{equation}}
\newcommand{\eneq}{\end{equation}}
\input{epsf}

\begin{document}

\title{Speeding up entanglement generation by proximity to higher-order exceptional points
}

\author{Zeng-Zhao Li}
\email{zengzhaoli09@gmail.com}
\affiliation{Department of Chemistry, University of California, Berkeley, California 94720, USA}
\affiliation{Berkeley Center for Quantum Information and Computation, Berkeley, California 94720, USA}

\author{Weijian Chen}
\email{wchen34@wustl.edu} 
\affiliation{Department of Physics, Washington University, St. Louis, Missouri 63130, USA}

\author{Maryam Abbasi}
\affiliation{Department of Physics, Washington University, St. Louis, Missouri 63130, USA}

\author{Kater W. Murch}
\affiliation{Department of Physics, Washington University, St. Louis, Missouri 63130, USA}

\author{K. Birgitta Whaley}
\email{whaley@berkeley.edu} 
\affiliation{Department of Chemistry, University of California, Berkeley, California 94720, USA}
\affiliation{Berkeley Center for Quantum Information and Computation, Berkeley, California 94720, USA}

\begin{abstract}
Entanglement is a key resource for quantum information technologies ranging from quantum sensing to quantum computing. Conventionally, the entanglement between two coupled qubits is established at the time scale of the inverse of the coupling strength. In this work, we study two weakly coupled non-Hermitian qubits and observe entanglement generation at a significantly shorter time scale by proximity to a higher-order exceptional point. We establish a non-Hermitian perturbation theory based on constructing a biorthogonal complete basis and further identify the optimal condition to obtain the maximally entangled state. Our study of speeding up entanglement generation in non-Hermitian quantum systems opens new avenues for harnessing coherent nonunitary dissipation for quantum technologies. 
\end{abstract}
\date{\today}
\maketitle

{\it Introduction}.---
Since the discovery of their real spectra, parity-time (${\cal PT}$) symmetric Hamiltonians \cite{Bender98prl,Bender07} a special class of non-Hermitian Hamiltonians have been intensively studied in the past two decades~\cite{El-GanainyChristodoulides18nphys,Ozdemir19nmat,AshidaGongUeda20}. One intriguing feature of such ${\cal PT}$-symmetric non-Hermitian systems is the nontrivial degeneracy known as the exceptional point (EP) where both eigenenergies and eigenstates coalesce.  This unique feature has been experimentally demonstrated in many classical systems~\cite{Dembowski01prl,HodaeiKhajavikhan14science,XuHarris16nature} with applications such as wave transport control~\cite{RuterKip10nphys,DopplerRotter16nature,ChoiBerini17ncomms,ZhangChan18prx}, laser emission management~\cite{BrandstetterRotter14ncomms,PengOzdemirYang16pnas,WongZhang16nphoton}, and enhanced sensing ~\cite{Wiersig14prl,ChenYang17nature,HodaeiKhajavikhan17nature}.
Recently, various  approaches such as dissipation engineering~\cite{NaghilooMurch19nphys} and Hamiltonian dilation~\cite{WuRongDu19science} have been utilized to study non-Hermitian physics in quantum systems~\cite{LewalleWhaley22,LiAtalayaWhaley22pra}. EPs have been observed in various quantum platforms such as the nitrogen-vacancy color center~\cite{WuRongDu19science}, trapped ions~\cite{DingZhang21prl,WangJingChen21pra}, ultracold atoms~\cite{LiJoglekarLuo19ncommun}, and superconducting circuits~\cite{NaghilooMurch19nphys},
and offer new possibilities in quantum applications such as sensing~\cite{YuLiGuo20prl} and state control~\cite{AbbasiChenMurch22prl,LiuDu21prl}. 
Although the majority of detailed studies focus on single spin or spin ensembles described in Hilbert space with dimension $N=2$, there are several pioneering works studying the coherence and entanglement in a larger Hilbert space~\cite{WenLong21prr,GautamDoraiArvind22arXiv}. 
A very recent work demonstrates entanglement between one effective non-Hermitian qubit and one Hermitian qubit that achieves the maximum allowed value at the EP~\cite{KumarMurchJoglekar22pra}.

In this Letter, we study the entanglement generation between two driven non-Hermitian qubits that are weakly coupled. In the absence of coupling, the two-qubit system exhibits a fourth-order EP. The weak coupling is a perturbation, lifting the degeneracy and altering the two-qubit dynamics. We observe a characteristic pattern of phase accumulation among the basis states, which lead to a maximally entangled state at a time scale much shorter than the inverse of the coupling strength. We further observe that by approaching the EP, a weaker coupling strength is needed to establish the maximally entangled state, at the cost of longer build-up time. In addition, we develop a non-Hermitian perturbation theory and use this to obtain analytical solutions of the concurrence evolution, which agrees well with the numerical simulations.

\begin{figure}
\centering
  \includegraphics[width=.99\columnwidth]{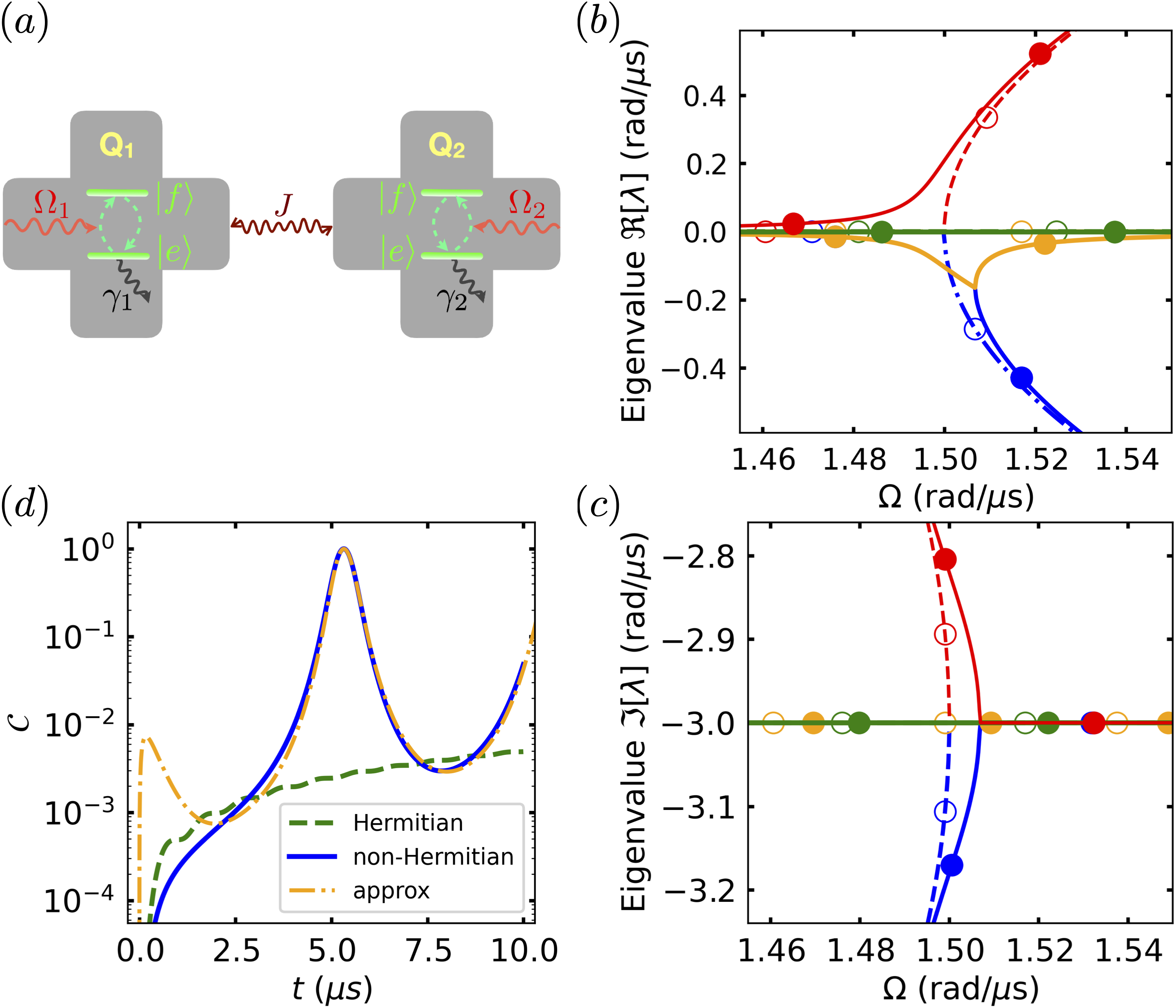} 
\caption{(color online) (a) Schematic of two coupled non-Hermitian qubits with coupling $J$. The lower level of each qubit has energy dissipation $\gamma_{j=1,2}$ and the two levels of each qubit are coupled by a drive $\Omega_{j=1,2}$. (b, c)  Real (b) and imaginary (c) parts of the eigenvalues for $J=0$ (dashed curves with empty circles) and for $J=0.001$ rad/$\mu$s (solid curves with full circles). The two qubits are assumed to have the same drive amplitude $\Omega$ and the same decay rates $\gamma_{1,2} = 6$ $\mu$s$^{-1}$. The circles are visual aids to distinguish the solid and dashed lines. (d) Concurrence evolution of the coupled non-Hermitian qubits near the fourth-order EP (solid blue curve). The yellow dashed curve shows the perturbative analytical result calculated with Eq.~(\ref{eq:C_approx})~\cite{SM}. The initial state is $|ff\rangle$ and other parameters used are: $\gamma_{1,2} = 6$ $\mu$s$^{-1}$, $\Omega_{1,2} = 1.6$ rad/$\mu$s, $\Delta_{1,2}=0$, and $J = 0.001\,\rm{rad}/\mu s$. The dashed green curve shows the concurrence evolution for two coupled Hermitian qubits with the same coupling  $J$, also with the initial state $|ff\rangle$. 
} 
\label{fig:Fig1_eigenvalues_nonHermi_vs_Hermi_approxEnergylevels_approxEnergylevels}
\end{figure}

{\it The model}.--- The system under consideration consists of two coupled driven non-Hermitian qubits, where the lower energy level has energy dissipation out of the qubit manifold, $\{|e\rangle, |f\rangle\}$, as shown in Fig.~\ref{fig:Fig1_eigenvalues_nonHermi_vs_Hermi_approxEnergylevels_approxEnergylevels}(a). Such a non-Hermitian qubit has been realized in a transmon superconducting circuit~\cite{NaghilooMurch19nphys}. 
The Hamiltonian (setting $\hbar=1$) for a single non-Hermitian qubit is given by
\begin{eqnarray}
H_{j=1,2} &=& (\Delta_j-\frac{i\gamma_j}{2}) \sigma_j^-\sigma_j^+ + \Omega_j\sigma_j^x ,
\label{eq:Hamilsingle}
\end{eqnarray}
where $\Delta_j$ represents frequency detuning of the applied drive from the qubit transition frequency,  $\gamma_j$ denotes the energy decay rate of $|e\rangle_j$, and $\Omega_j$ is the drive amplitude. 
The Pauli operators are defined in terms of energy levels $|e\rangle_j$ and $|f\rangle_j$ as $\sigma_j^+ = |f\rangle_j\langle e|$, $\sigma_j^- = |e\rangle_j\langle f|$, and $\sigma_j^x=
|f\rangle_j\langle e| +|e\rangle_j\langle f|$ ($j=1, 2$). The eigenvalues of $H_j(\Delta_j = 0)$ are given by $\lambda_{j,\pm} = (-i\gamma_j \pm \eta_j)/4$, with $\eta_j \equiv \sqrt{16 \Omega_j^2 - \gamma_j^2}$, and there exists a second-order EP when $\Omega_j = \gamma_j/4$.

The coupled non-Hermitian qubits are then described by the Hamiltonian 
\begin{eqnarray}
H &=& \sum_{j=1,2} H_j + J (\sigma_1^+\sigma_2^- + \sigma_1^-\sigma_2^+) ,
\label{eq:Hamil}
\end{eqnarray}  
where $J$ denotes the effective coupling strength between the two qubits. 
When the drives are resonant with both qubits, i.e., $\Delta_j=0$, Eq.~(\ref{eq:Hamil}) reduces to a so-called passive ${\cal PT}$-symmetric Hamiltonian~\cite{GuoChristodoulides09prl} which can be further written as $H(\Delta_j=0)=H_{\cal PT}-\frac{i\gamma_1}{4}-\frac{i\gamma_2}{4}$, where the Hamiltonian $H_{\cal PT}$ respects ${\cal PT}$-symmetry, i.e., ${\cal PT}H_{\cal PT}({\cal PT})^{-1}=H_{\cal PT}$ with parity operator ${\cal P}=\sigma_1^x\sigma_2^x$ and ${\cal T}$ the time reversal operator (equivalent to complex conjugation).

We restrict our attention to the resonant case and assume the Hamiltonians for the two qubits are identical, i.e., $\Delta_1=\Delta_2=0$, $\gamma_1=\gamma_2=\gamma$, and $\Omega_1=\Omega_2=\Omega$. In the absence of qubit coupling (i.e., $J=0$), each qubit in its Hilbert space with dimension $N=2$ exhibits a second-order EP, while the two-qubit system described by the Hamiltonian in Eq.~(\ref{eq:Hamil}) with a Hilbert space with dimension $N=4$ exhibits a fourth-order EP at $\Omega=\Omega_{\rm EP}\equiv\frac{\gamma}{4}$ [see 
Figs.~\ref{fig:Fig1_eigenvalues_nonHermi_vs_Hermi_approxEnergylevels_approxEnergylevels}(b) and \ref{fig:Fig1_eigenvalues_nonHermi_vs_Hermi_approxEnergylevels_approxEnergylevels}(c) and 
Fig.~S1(a) in~\cite{SM}]. 
The weak coupling $J$ between the two qubits acts as a perturbation, lifting the degeneracy and lowering the order of the EP [e.g., 
to a second-order EP at a shifted position 
in Figs.~\ref{fig:Fig1_eigenvalues_nonHermi_vs_Hermi_approxEnergylevels_approxEnergylevels}(b) and \ref{fig:Fig1_eigenvalues_nonHermi_vs_Hermi_approxEnergylevels_approxEnergylevels}(c)]\cite{SM}.

The two-qubit evolution can be solved from the Hamiltonian in Eq.~(\ref{eq:Hamil}), and the normalized quantum state has a general form $|\tilde{\psi}\rangle=\frac{|\psi\rangle}{||\psi\rangle|} =\alpha|ff\rangle+\beta|fe\rangle+\zeta|ef\rangle+\delta|ee\rangle$, where the 
state $|\psi\rangle$ is normalized conventionally. 
We use the concurrence $\cal C$~\cite{BennettSchumacher96pra} to quantify the entanglement between the two qubits, which is given by
${\cal C}= 2|\alpha\delta-\beta\zeta|$. 
Conventionally, two qubits with dipolar coupling as in Eq.~(\ref{eq:Hamil}) and initialized in the state $|fe\rangle$ (or $|ef\rangle$) will generate Bell states on a time scale of $1/J$. Here we find that by ensuring proximity to the fourth-order EP, weakly coupled non-Hermitian qubits ($J \ll \gamma$) with initial state $|ff\rangle$ can have entanglement generation on a time scale much shorter than $1/J$. 
Figure~\ref{fig:Fig1_eigenvalues_nonHermi_vs_Hermi_approxEnergylevels_approxEnergylevels}(d) shows an example of this EP-enhanced entanglement generation, together with a comparison to the Hermitian case.

\begin{figure*}
\centering
  \includegraphics[width=2\columnwidth]{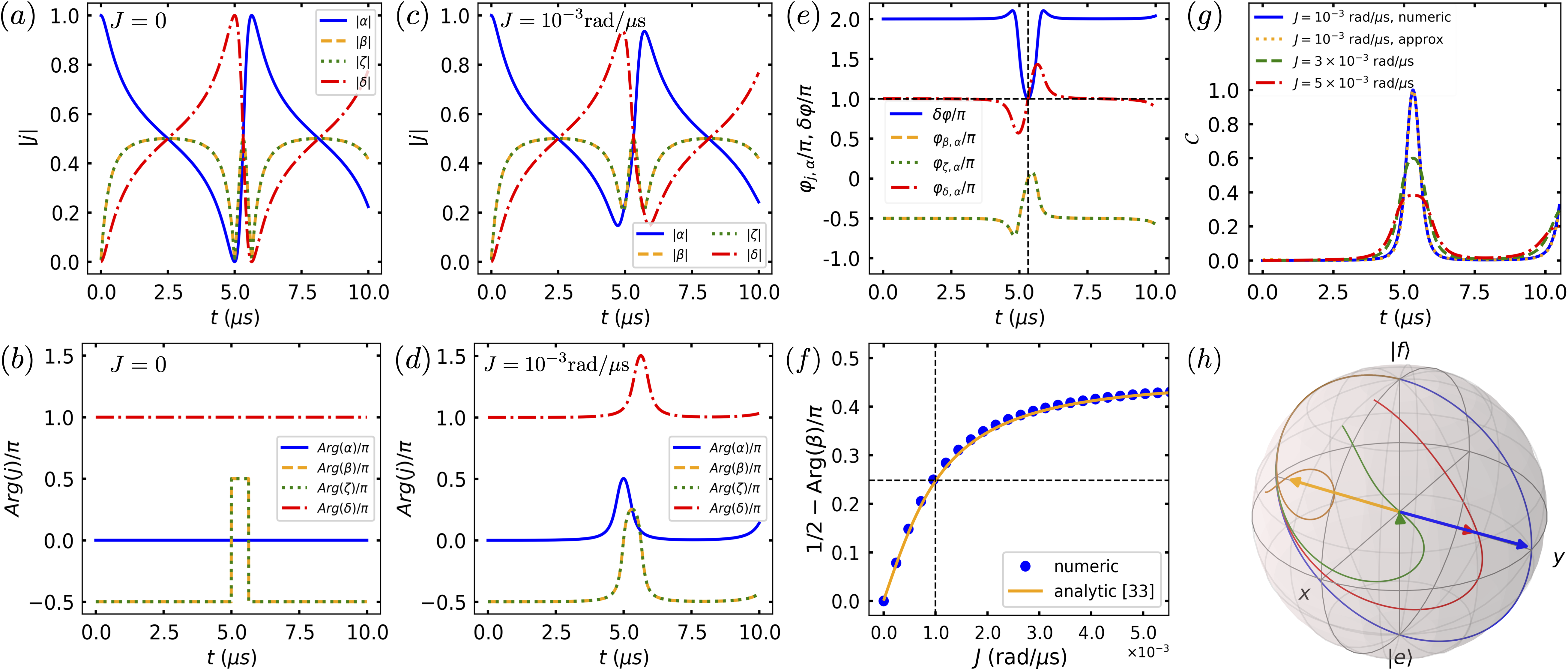} 
\caption{(color online) (a-d) Time evolution of the modulus $|j|$ and the angle ${\rm Arg}(j)$ ($j=\alpha,\beta,\zeta,\delta$) for each of the four complex amplitudes of the two-qubit state $|\tilde{\psi}\rangle =\alpha|ff\rangle+\beta|fe\rangle+\zeta|ef\rangle+\delta|ee\rangle$ for $J=0$ (a, b) and $J=10^{-3}$ rad/$\mu$s (c, d). 
(e) Time evolution of the phase relative to basis state $|ff\rangle$, i.e., $\varphi_{j,\alpha}\equiv{\rm Arg}(j)-{\rm Arg}(\alpha)$ for ($j=\beta, \zeta, \delta$) and of $\delta\varphi\equiv{\rm Arg}(\alpha)+{\rm Arg}(\delta) -{\rm Arg}(\beta)-{\rm Arg}(\zeta)$ for $J=10^{-3}$ rad/$\mu$s. The value $\delta\varphi= \pi$ at $t=T_0^* =5.325$ $\mu$s signifies maximal entanglement. (f) Dependence on $J$ of ${\rm Arg}(\beta)$  relative  to its value for $J=0$ (panel (b)), i.e., the differential phase $\pi/2- {\rm Arg}(\beta)$, at time $t=5.325$ $\mu$s. 
(g) Examples of concurrence for $J=10^{-3},\,3\times10^{-3},\,5\times10^{-3}$ rad/$\mu$s. Maximal concurrence generation (${\cal C}=1$) is observed at $t= 5.325\,\mu$s for $J=10^{-3}$ rad/$\mu$s, with the values from Eq.~(\ref{eq:C_approx}) (orange dotted curve) in good agreement with the full numerical calculation (blue solid curve). 
(h) Trajectories of reduced single qubit dynamics on the reduced qubit Bloch sphere for $J=0$ (blue), $5\times 10^{-4}$ (red), $10^{-3}$ (green), $5\times 10^{-3}$ rad/$\mu$s (orange) with evolution time between 0 and $t=4\pi/\eta = 5.64\,\mu$s (i.e., a single period for decoupled qubits with $J=0$). Arrows point to the reduced qubit states at time $t= 5.325$ $\mu$s, with the blue arrow pointing to $|+y\rangle$, the green arrow pointing to the origin (indicating a maximally entangled state), and the red and orange arrows aligning with the $\pm y$ axis. Unless otherwise specified, all plots are made for $|\psi(0)\rangle=|ff\rangle$, $\Omega=1.6$ rad/$\mu$s, and $\gamma=6$ $\mu$s$^{-1}$.}  
\label{fig:Fig2_EP3_enhancement}
\end{figure*}

{\it Non-Hermitian perturbation theory}.---We construct a biorthogonal complete basis from the states in both the Hilbert and corresponding dual spaces of the two qubits and extend the standard perturbation theory of Hermitian quantum mechanics to develop a time-independent perturbation theory for non-Hermitian systems~\cite{SM}. In addition, we also generalize standard degenerate perturbation theory to deal with the existence of non-Hermitian degeneracies~\cite{SM}.

With perturbative eigenstates $|\Psi_{j}\rangle$ and eigenvalues $\Lambda_{j}$ for 
$\Omega > \Omega_{\rm EP}$~\cite{SM}, 
we can obtain the state evolution $|\tilde{\psi}(t)\rangle = \frac{|\psi(t)\rangle}{| |\psi(t)\rangle |} $, where
$|\psi(t)\rangle = \sum_{j=++,--,1,2}\langle \bar{\Psi}_{j} |\psi(0)\rangle e^{-it\Lambda_{j}} | \Psi_{j}\rangle  $,
with $\langle \bar{\Psi}_{++}|, \langle \bar{\Psi}_{--}|, \langle \bar{\Psi}_{1}|, \langle \bar{\Psi}_{2}|$ the corresponding left eigenstates in the biorthogonal basis. 
The use of a biorthogonal basis is crucial here since the Hamiltonian Eq.~(\ref{eq:Hamil}) is not in general ${\cal PT}$-symmetric, so the usual ${\cal CPT}$ norm for ${\cal PT}$-symmetric Hamiltonians~\cite{Bender02prl,Bhosale21} does not apply. 
For the initial state $|\psi(0)\rangle=|ff\rangle$, the time evolution of concurrence is then given by~\cite{SM} 
\begin{eqnarray}
{\cal C} &=& \Big| 
\frac{16 \eta^2 \Omega^2 \left(e^{it\chi_2} -1 \right)}{\cal B }
\Big| ,
\label{eq:C_approx}
\end{eqnarray}
where ${\cal B} = 16\Omega^2(\gamma^2+32\Omega^2) +\gamma \Big\{\gamma(\gamma^2 -\eta^2)\cos(t\eta) -2\gamma^2\eta\sin(t\eta) 
-64\Omega^2\cos(\frac{t\chi_2}{2}) \Big[\gamma\cos\left(\frac{t\eta}{2}\right) - \eta\sin\left(\frac{t\eta}{2}\right)\Big] \Big\} $ and $\chi_2 = J\frac{\eta^2 + 3\gamma^2}{\eta^2}$. 
The analytical result from Eq.~(\ref{eq:C_approx}) agrees well with the numerical simulation results in Fig.~\ref{fig:Fig1_eigenvalues_nonHermi_vs_Hermi_approxEnergylevels_approxEnergylevels}(d) (except for the region close to $t=0$~\cite{SM}). 
In the following sections we discuss the mechanism of entanglement generation in detail and show how the entanglement generation is enhanced by approaching the fourth-order EP.

{\it Entanglement generation near the EP}.---We first examine the complex amplitudes of $|\tilde{\psi}(t)\rangle$ at each basis state in the absence of the qubit coupling, i.e., $J=0$ [Figs.~\ref{fig:Fig2_EP3_enhancement}(a) and \ref{fig:Fig2_EP3_enhancement}(b)]. The qubit drive is chosen as $\Omega = 1.6$ rad/$\mu$s, which places the system in the regime of ${\cal PT}$-symmetry preserving phase. The two qubits evolve independently; therefore the two-qubit state 
evolves successively through states $|ff\rangle$, $(|f\rangle-i|e\rangle)\otimes (|f\rangle-i|e\rangle)$, $|ee\rangle$, and $(|f\rangle+i|e\rangle)\otimes (|f\rangle+i|e\rangle)$, returning to $|ff\rangle$ after one period (i.e., $t=4\pi/\eta\sim5.64\,\mu$s).  
An interesting feature is the distorted Rabi-like oscillation of the population of $|ee\rangle$ and $|ff\rangle$ states, shown as the blue and red curves in Fig.~\ref{fig:Fig2_EP3_enhancement}(a). 
Inspection of the behavior shortly before $~5.64$ $ \mu$s reveals  that at the time $t=T^*_0=5.325$ $\mu$s all basis states have equal amplitude.
Fig.~\ref{fig:Fig2_EP3_enhancement}(b) shows that the phases of the $|ef\rangle$ and $|fe\rangle$ states experience sudden jumps of $\pi$, corresponding to the state of one of the two qubits passing through a pole of the corresponding qubit Bloch sphere. 
When all basis amplitudes are equal, the concurrence, ${\cal C}= 2|\alpha\delta-\beta\zeta|$, can be shown to be equal to $|\sin(\frac{\delta\varphi}{2})|$ with $\delta\varphi\equiv {\rm Arg}(\alpha) +{\rm Arg}(\delta) -{\rm Arg}(\beta) -{\rm Arg}(\zeta)$. 
Evaluating the phase $\delta\varphi$ at $T^*_0$ it is evident that concurrence $\cal{C}$ is identically equal to zero. 
This is consistent with evaluation of Eq.~(\ref{eq:C_approx}) with $J=0$.

Figures~\ref{fig:Fig2_EP3_enhancement}(c) and \ref{fig:Fig2_EP3_enhancement}(d) now summarize the state evolution with a finite but weak coupling strength ($J=10^{-3}$ rad/$\mu$s). The populations of the basis states in Fig.~\ref{fig:Fig2_EP3_enhancement}(c) exhibit a similar distorted oscillatory behavior, but with reduced visibility relative to that for $J=0$. Here the two-qubit state does not return to the product states $|ff\rangle$ and $|ee\rangle$ because of the qubit coupling. Also, while at the time $T^*_0$ we still observe that $|\alpha|\sim|\beta|\sim|\zeta|\sim|\delta|$, Fig.~\ref{fig:Fig2_EP3_enhancement}(d) shows that the phase evolution of each basis state is now significantly altered near $T^*_0$. 
In particular, the $\pi$ phase jumps of ${\rm Arg}(\beta)$ and ${\rm Arg}(\zeta)$ that are seen for $J=0$ in Fig.~\ref{fig:Fig2_EP3_enhancement}(b), now become Gaussian-shape profiles with significantly reduced phase contrast. 
Physically speaking, we can intuitively understand this effect as a result of the weak qubit-qubit coupling $J$ pulling the qubit dynamics away from the poles of the Bloch sphere, so that no discrete $\pi$ phase jumps occur. 

The time evolution of the relative phases with respect to that of basis state $|ff\rangle$, i.e., $\varphi_{j,\alpha}\equiv {\rm Arg}(j) -{\rm Arg}(\alpha)$ ($j=\beta, \zeta, \delta$), is shown for coupling strength $J=10^{-3}$ rad/$\mu$s in Fig.~\ref{fig:Fig2_EP3_enhancement}(e), together with the phase $\delta\varphi$ (solid blue line). 
At $T_0^*$ where the populations at each basis state are still approximately equal, the relative phase $\varphi_{\delta,\alpha}$ is still equal to $\pi$, as in Fig.~\ref{fig:Fig2_EP3_enhancement}(b) for $J=0$, but the values of $\varphi_{\beta,\alpha}$  and $\varphi_{\zeta,\alpha}$ are now no longer equal to $\pi/2$. 
In fact, 
the state is very close to $|\tilde{\psi}\rangle = e^{i\frac{\pi}{4}}( |ff\rangle + |fe\rangle + |ef\rangle + e^{+i\pi}|ee\rangle)/2$ which has $\delta\varphi = \pi$ and ${\cal C}=1$.

Fig.~\ref{fig:Fig2_EP3_enhancement}(f) now shows the change in phase of ${\rm Arg}(\beta)$ relative to its value at $J=0$, i.e., the differential phase $\pi/2 - {\rm Arg}(\beta)$, as a function of the coupling strength $J$.
The analytical results (Eq. (S77) in \cite{SM}) agree well with the full numerical results.  At a given drive amplitude $\Omega$, there exists an optimal coupling strength $J$ to satisfy the condition $\delta \varphi = \pi$ that realizes the maximally entangled state on a time scale $1/\eta$ $(\ll 1/J)$. 
This change in differential phase of $|fe\rangle$ and $|ef\rangle$ can be understood as a consequence of the degeneracy lifting by a finite $J$ value (with $J \ll \Omega$). 
Fig.~\ref{fig:Fig2_EP3_enhancement}(g) shows that further increase of $J$ does not benefit the entanglement generation. The analytical expression from Eq.~(\ref{eq:C_approx}) for $J=10^{-3}$ rad/$\mu$s shows good agreement with the numerical simulations.

\begin{figure}
\centering
  \includegraphics[width=0.86\columnwidth]{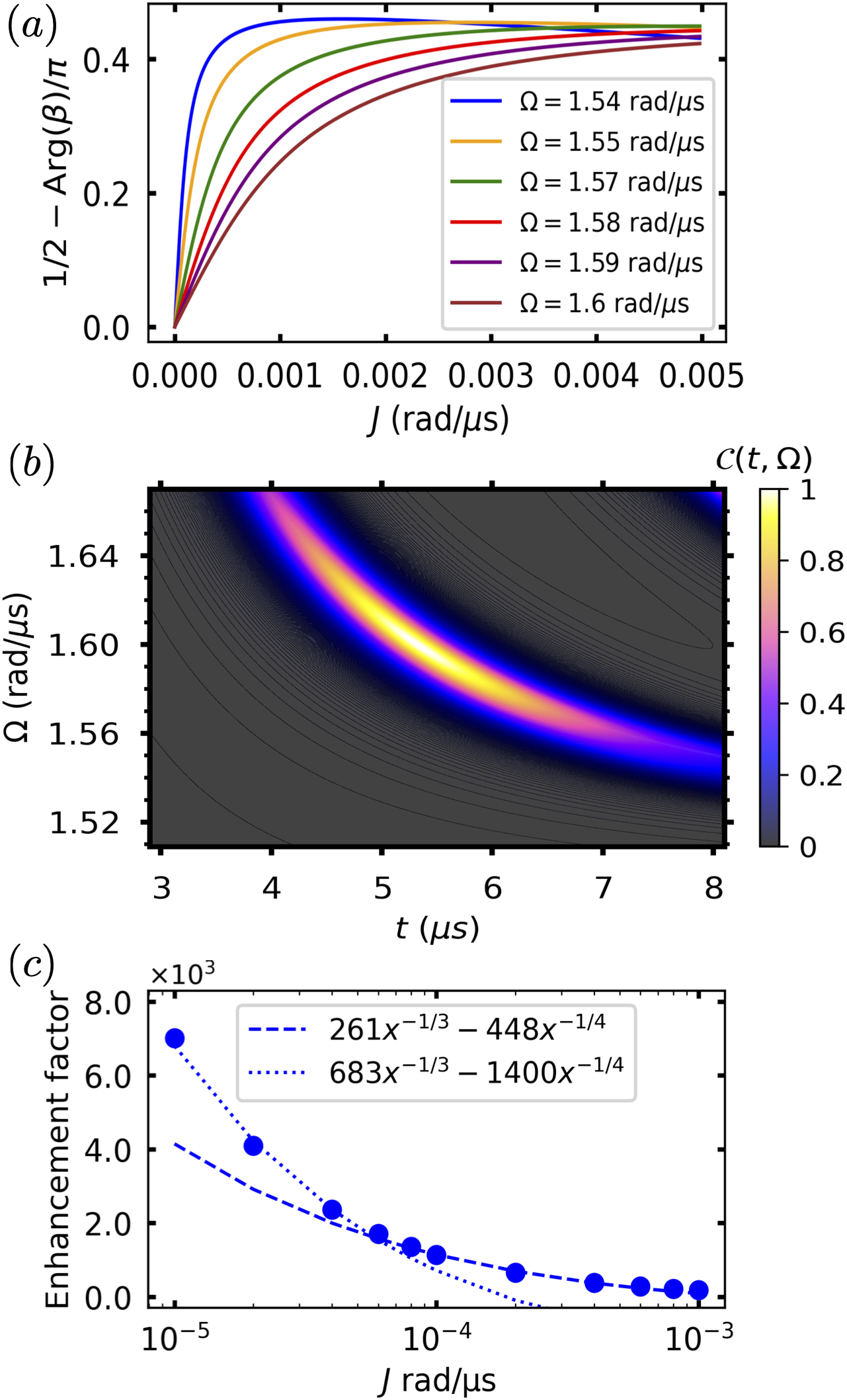} 
\caption{(color online) (a) Phase of $\beta$ relative to the corresponding value for $J=0$, i.e., $\pi/2-{\rm Arg}(\beta)$, at different $\Omega$ values. 
The increased slope on approaching the EP at $\Omega_{\rm EP}=1.5$ rad/$\mu$s indicates the enhancement due to proximity to the EP.
(b) Density plot of concurrence as a function of drive amplitude $\Omega$ and time $t$, with $J$ fixed at $10^{-3}$ rad/$\mu$s. Identification of the maximal concurrence ${\cal C}=1$ yields the optimal parameters $(\Omega^*, T^*) = (1.6\,{\rm rad}/\mu\rm{s}, 5.325\,\mu\rm{s})$. 
(c) Entanglement enhancement factor defined as the ratio of the first time  to achieve maximal entanglement for the Hermitian case ($\sim 1/J$) to the corresponding time $T^*$ for the non-Hermitian case 
(see also Fig.S2 in~\cite{SM}). 
}
\label{fig:Fig3_optimalOmega}
\end{figure}

To get more physical intuition into this enhancement of entanglement, 
we also calculate the reduced single qubit dynamics, i.e., after tracing out one of the two qubits. Figure~\ref{fig:Fig2_EP3_enhancement}(h) shows the trajectories of the reduced qubit dynamics on the corresponding Bloch sphere during one period $4\pi/\eta = 5.64\,\mu$s. For $J=0$, the reduced qubit evolves on the surface of the Bloch sphere, while for non-zero $J$, the reduced qubit dynamics lie inside the Bloch sphere, suggesting possible entanglement generation. Only at a specific value of $J$ will the qubit trajectory pass through the origin where the reduced qubit is in a fully mixed state, implying that two qubits are in a maximally entangled state. It is then evident that for a given value of $\Omega$, maximum entanglement will occur at a specific time $T^*$ 
and for a specific coupling strength $J^*$.

{\it EP enhanced entanglement generation}.--- Next, we study the differential phase of the $|ef\rangle$  and $|fe\rangle$ states relative to their $J=0$ values, i.e., $\pi/2-{\rm Arg}(\beta)$ at $T^{*}$, for different values of the drive amplitude $\Omega$ [Fig.~\ref{fig:Fig3_optimalOmega}(a)]. On approaching the EP, the slope of the phase change gets sharper, and therefore a smaller coupling strength is needed to achieve the maximally entangled state. 
The increasingly sharp change in differential phase on approaching the EP [Fig.~\ref{fig:Fig3_optimalOmega}(a)] suggests that the entanglement generation near the EP can provide a sensitive measure of the magnitude of the qubit coupling $J$~\cite{Wiersig20}. 
%
The density plot of concurrence versus $\Omega$ and $t$ in Fig.~\ref{fig:Fig3_optimalOmega}(b) shows how the optimal parameters $\{\Omega^*, T^*\}=\{1.6\,{\rm rad/\mu s}, 5.325\,\mu {\rm s}\}$ can be identified for a given qubit coupling $J$. For this example with $J=10^{-3}$ rad/$\mu$s, the maximal concurrence ${\cal C}=1$ corresponds to the peak of the blue curve in Fig.~\ref{fig:Fig2_EP3_enhancement}(g). 
Fig.~\ref{fig:Fig3_optimalOmega}(c) shows the entanglement enhancement for non-Hermitian qubits relative to their corresponding Hermitian qubits. It is evident that non-Hermitian qubits can be entangled significantly faster than Hermitian qubits, with the relative speedup increasing significantly as the perturbation $J$ is decreased and the fourth-order EP at $J=0$ is approached.
The mixed power-law dependence, i.e., a combination of inverse cube root and fourth root, reflects the changing order of the EP as $J$ decreases. 

\begin{figure}
\centering
  \includegraphics[width=0.8\columnwidth]{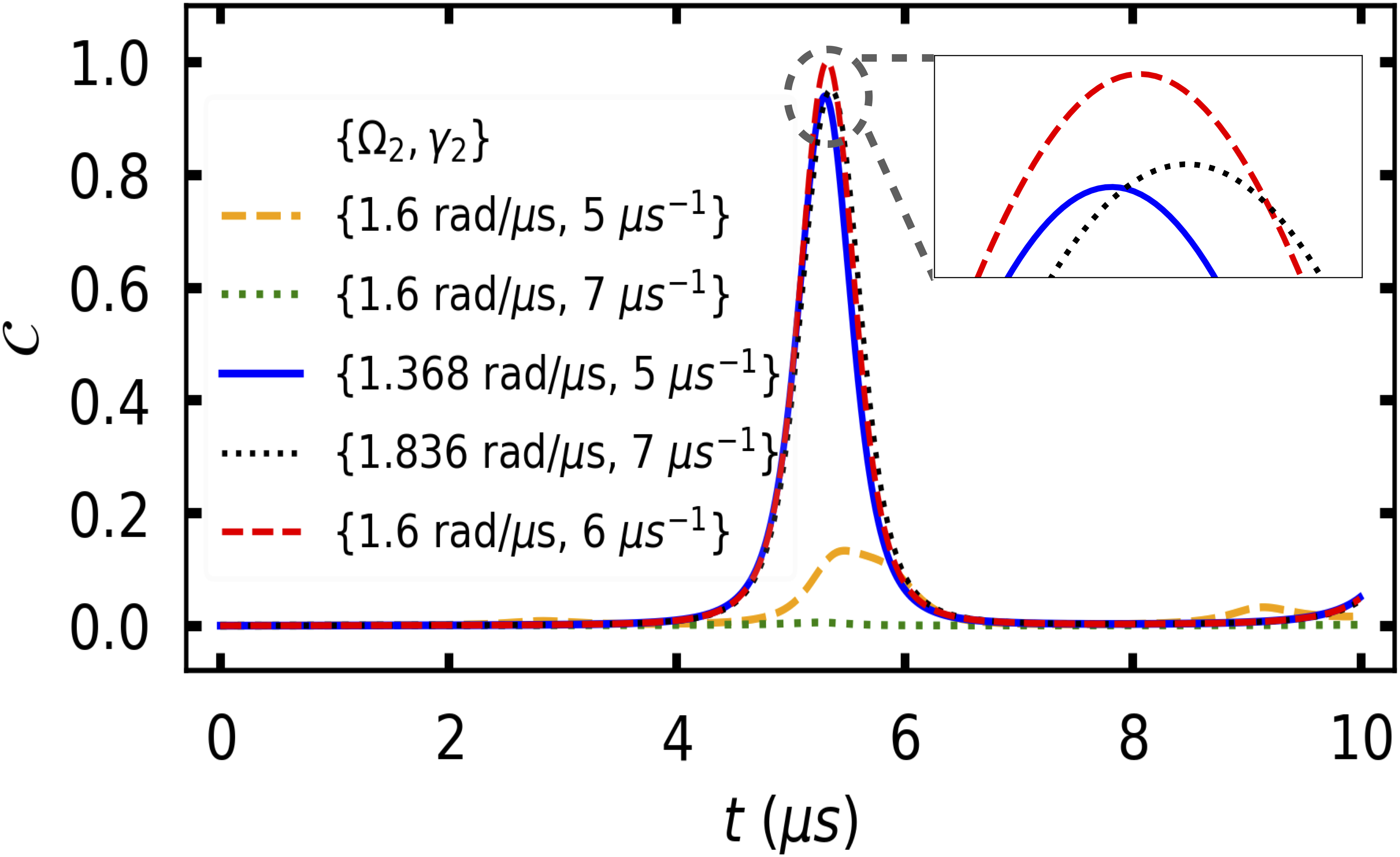} 
\caption{(color online) Time evolution of concurrence for two qubits with different Hamiltonian parameters. $\{\Omega_1, \gamma_1\}=\{1.6\,{\rm rad}/\mu{\rm s}, 6\,\mu{\rm s}^{-1}\}$ are fixed, while the parameters of the second qubit take on the values shown in the legend. The red dashed curve for $\{\Omega_2, \gamma_2\} = \{\Omega_1, \gamma_1\}$ is included as a reference. The initial state and coupling strength are given by $|\psi(0)\rangle=|ff\rangle$ and $J=10^{-3}$ rad/$\mu$s, respectively, for all calculations.} 
\label{fig:Fig4_nonHermitian_return}
\end{figure}

{\it Discussion}.---Throughout this work, we have focused on the case of two non-Hermitian qubits with identical Hamiltonian parameters, which may be challenging to realize in experiments. However, our mechanism is robust with respect to the parameter values and still holds when the qubit parameters differ within a small range. In Fig~\ref{fig:Fig4_nonHermitian_return}, we compare the concurrence evolution for fixed 
$\gamma_1 = 6\,\mu \rm{s}^{-1}$ and variable $\gamma_2 = 5, 7\,\mu \rm{s}^{-1}$. 
The concurrence degrades significantly if both $\Omega_1$ and $\Omega_2$ are maintained at $1.6\,\mathrm{rad}/\mu$s, as chosen in Fig.~\ref{fig:Fig2_EP3_enhancement}. 
However, if $\Omega_2$ is adjusted to ensure the two qubits have the same period, i.e., $16\Omega_1^2-\gamma_1^2=16\Omega_2^2-\gamma_2^2$, an entangled state with concurrence ${\cal C} \approx 1$ is regained.   
We further note that our method still provides an advantage for relatively large $J$ values. For example, for $J = 0.1$ rad/$\mu$s, the maximally entangled state can be obtained in less than $2$ $\mu$s, while it takes about $8$ $\mu$s for Hermitian qubits to reach this state (see Fig.~S4 in~\cite{SM}). 
%
More detailed discussion on the effects of a non-resonant qubit drive, of decoherence, and of the entanglement generation in the Hermitian limit is given in~\cite{SM}.

{\it Conclusions}.---We have studied the entanglement between two weakly coupled non-Hermitian qubits that can be generated at a speed much faster than the inverse of the inter-qubit interaction, by forcing proximity to a higher-order exceptional point. 
This demonstrates a fundamentally new and advantageous entanglement generation method, which offers considerable potential for applications in quantum information science and technology, including the sensing of weak coupling between qubits and common two-level system impurities or defects.

\acknowledgements
This work has been supported by AFOSR MURI Grant No. FA9550-21-1-0202. We are grateful to Philippe Lewalle and Rob Cook for helpful discussions.

Z.-Z.L. and W.C. contributed equally to this work.

\end{document}


\title{Supplemental materials for ``Speeding up entanglement generation by proximity to higher-order exceptional points''}

\author{Zeng-Zhao Li}
\email{zengzhaoli09@gmail.com}
\affiliation{Department of Chemistry, University of California, Berkeley, California 94720, USA}
\affiliation{Berkeley Center for Quantum Information and Computation, Berkeley, California 94720, USA}

\author{Weijian Chen}
\email{wchen34@wustl.edu} 
\affiliation{Department of Physics, Washington University, St. Louis, Missouri 63130, USA}

\author{Maryam Abbasi}
\affiliation{Department of Physics, Washington University, St. Louis, Missouri 63130, USA}

\author{Kater W. Murch}
\affiliation{Department of Physics, Washington University, St. Louis, Missouri 63130, USA}

\author{K. Birgitta Whaley}
\email{whaley@berkeley.edu} 
\affiliation{Department of Chemistry, University of California, Berkeley, California 94720, USA}
\affiliation{Berkeley Center for Quantum Information and Computation, Berkeley, California 94720, USA}

\begin{abstract}
In this supplemental material, we provide supplementary figures to support our main conclusion and a detailed derivation of the analytical expression of concurrence [i.e., Eq.~(3) in main text] by developing a first-order perturbation theory for two weakly coupled non-Hermitian qubits. 
In addition, we also discuss the effect of nonresonant qubit drive and decoherence as well as entanglement generation in the Hermitian limit. 
\end{abstract}
\date{\today}
\maketitle
\tableofcontents

\section{Additional figures \label{app:figures}}

\begin{figure}[htp]
\centering
  \includegraphics[width=0.9\columnwidth]{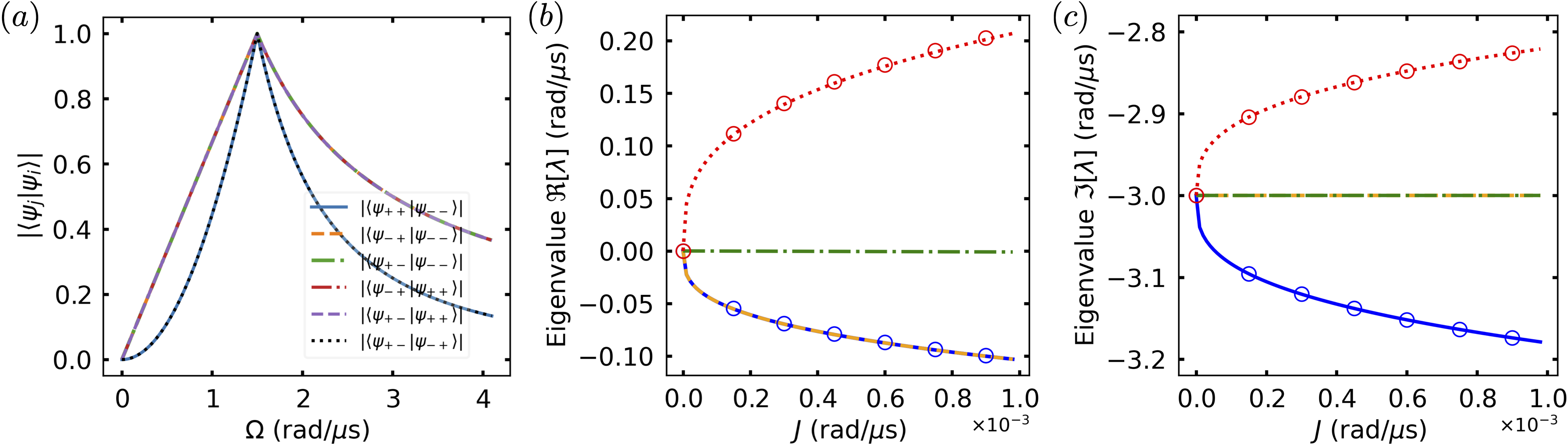} 
\caption{{\bf Demonstration of the fourth-order EP.} (a) Absolute value of the overlap of the eigenvectors $|\psi_i\rangle$. 
The four eigenvectors $|\psi_{--}\rangle$, $|\psi_{++}\rangle$, $|\psi_{-+}\rangle$, and $|\psi_{+-}\rangle$ have mutual overlap equal to 1 and thus coalesce at the fourth-order EP with $\Omega_{\rm EP}=1.5$ rad/$\mu$s. The parameters used are: $\gamma=6\,\mu\rm{s}^{-1}$ and $J=0$. 
(b, c) The cube-root dependence of the real (b) and imaginary (c) parts of
the eigenvalues degenerate at $J=0$ 
on the qubit coupling $J$ when $\Omega=1.5$ rad/$\mu$s (dashed, dotted, dashed-dotted, or solid curves). Real parts of eigenvalues in (b) [$\Re\{\lambda_1\}$ (red), $\Re\{\lambda_2\}$ (green),  $\Re\{\lambda_3\}$ (orange), $\Re\{\lambda_4\}$ (blue)] are shown in descending order 
and the corresponding imaginary parts are shown in (c). In (b), the red and blue circles are calculated from $2.1\sqrt[3]{x}$ and $-1.03\sqrt[3]{x}$, respectively; in (c), the red and blue circles are calculated from $-3\pm\sqrt[3]{x}$. Here, $\gamma=6\,\mu\rm{s}^{-1}$. 
As finite perturbation lowers the order of the EP (e.g., the third order at $J=10^{-8}$ rad/$\mu$s with $\Omega=1.5$ rad/$\mu$s), it is expected that the resulting energy differences or transition frequencies will follow the cube-root relation. 
}  
\label{fig:Overlap_eigenstates}
\end{figure}

The fourth-order EP for $J=0$ is located at $\eta \equiv \sqrt{16\Omega^2 -\gamma^2}=0$. 
Confirmation of the fourth-order nature of this EP 
can be made by calculating the overlap of eigenvectors. 
Figure~\ref{fig:Overlap_eigenstates}(a) shows that for $J=0$, with $\gamma=6\,\mu\rm{s}^{-1}$, the four eigenvectors $|\psi_{--}\rangle$, $|\psi_{++}\rangle$, $|\psi_{-+}\rangle$, and $|\psi_{+-}\rangle$ coalesce at $\Omega = 1.5$ rad/$\mu$s (satisfying $\eta=0$),
where they all have unit overlap. 
The overlap between the eigenvectors was calculated for $\eta>0$ values from the analytical expressions for eigenvectors in the ${\cal PT}$-symmetric preserving phase 
given in Sec.~\ref{app:unbrokenPhase}
and for $\eta<0$ values  by the analytical expressions for the eigenvectors in the 
${\cal PT}$-symmetric broken phase 
given in Sec.~\ref{app:brokenPhase}. The eigenvectors at the EP 
are given by the $\eta=0$ limits of these expressions. 
%

Any finite coupling $J$ between the two qubits lifts the $J=0$ degeneracy and lowers the order of the fourth-order EP first to third order (e.g., $J=10^{-8}$ rad/µs with $\Omega=1.5$ rad/$\mu$s) and then for larger $J$ values, further down to second order ($J=10^{-3}$ rad/$\mu$s with $\Omega=1.506$ rad/$\mu$s, see the intersection of solid yellow and blue lines in
Figs.~1(b) and 1(c) of the main text). 
For much larger $J$ values there will be no EP, 
since as $J$ becomes large enough to be dominant over $\Omega$ and $\gamma$, the system will approach the Hermitian limit and there are no EPs.
Thus in general, a larger value of $J$ provides a greater reduction in the order of any EP in the system. 

We further examine the dependence of the real and imaginary parts of the eigenvalues on the qubit coupling strength $J$ when $\Omega = 1.5$ rad/$\mu$s. Figs.~\ref{fig:Overlap_eigenstates}(b) and \ref{fig:Overlap_eigenstates}(c) show that these follow a cube-root relation unless they are constant. 
The eigenvalues were calculated here by numerical diagonalization of the total Hamiltonian in the computational basis. 
The orange curve in Fig.~\ref{fig:Overlap_eigenstates}(b) shows a cube-root response of the real part to the perturbation while the corresponding imaginary part in Fig.~\ref{fig:Overlap_eigenstates}(b) is constant. The non-zero imaginary component (orange or green curve in Fig.~\ref{fig:Overlap_eigenstates}(c)) derives essentially from the passive ${\cal PT}$-symmetric Hamiltonian (equivalent to the ${\cal PT}$-symmetric part with balanced gain and loss accompanied by an imaginary offset, as demonstrated in the main text) that is under consideration in this work.

\begin{figure}[htp]
\centering
\includegraphics[width=0.5\columnwidth]{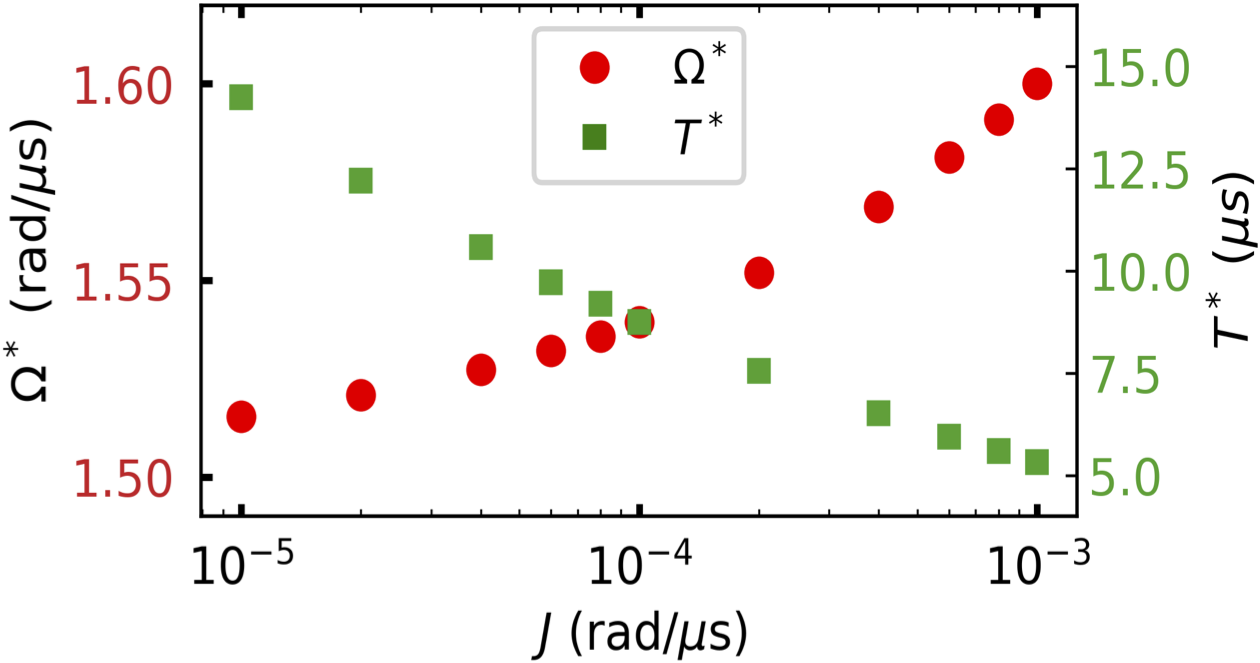} 
\caption{{\bf Optimal enhancement generation.} Dependence of $\Omega^*$ (red circles) and $T^*$ (green squares) on qubit coupling strength $J$. All calculations employ $|\psi(t=0)\rangle=|ff\rangle$ and $\gamma=6$ $\mu$s$^{-1}$.   
} 
\label{fig:FigS_optimal_Omega_T}
\end{figure}

Figure~\ref{fig:FigS_optimal_Omega_T} shows the optimal parameter combinations $\{\Omega^*, T^*\}$  for a broad range of values of coupling strength $J$, in addition to the parameter combination illustrated in Fig.~3(b) in the main text. It is evident that the weaker the coupling strength $J$, the closer the optimal $\Omega^*$ is to the EP value, at the cost however of a longer time duration required to establish maximal entanglement. Nevertheless, the time duration is always less than that required to generate maximal entanglement for the Hermitian system with the same $J$ value, ensuring enhancement of entanglement for all $J$ values (see Fig. 3(c) of the main text). When $J\rightarrow0$, it is expected that $\Omega^*\rightarrow\Omega_{\rm EP}$ and $T^*\rightarrow\infty$.

\begin{figure}[htp]
\centering
\includegraphics[width=0.36\columnwidth]{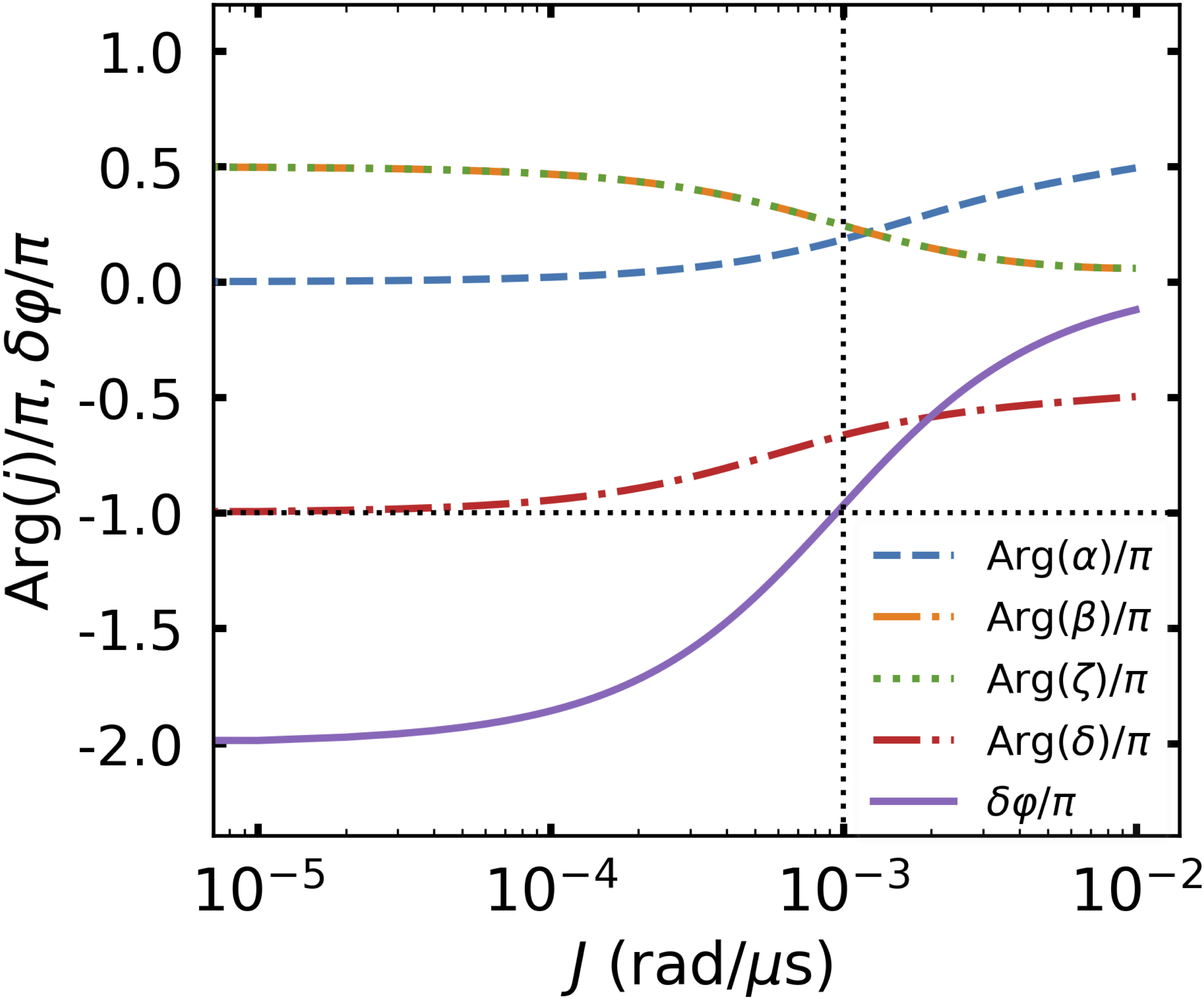} 
\caption{{\bf Dependence of the phases ${\rm Arg}(j)$ ($j=\alpha, \beta, \zeta, \delta$) of the complex coefficients on the coupling strength $J$.} The evolution time is set to $t=5.325$ $\mu$s at which time the populations of the basis states are approximately equal. The dependence of $\delta\varphi\equiv{\rm Arg}(\alpha)+{\rm Arg}(\delta) -{\rm Arg}(\beta)-{\rm Arg}(\zeta)$ (solid purple line) on the coupling strength is also shown. Maximal entanglement occurs at $\delta\varphi=-\pi$ [with a period $2\pi$ and consistent with $\pi$ in Fig.~2(e) in the main text], with the corresponding coupling $J=10^{-3}$ rad/$\mu$s, marked by the vertical and horizontal dotted lines in black. The initial state  $|\psi(0)\rangle=|ff\rangle$, and other parameters used are: $\gamma=6\,\mu{\rm s}^{-1}$ and $\Omega = 1.6\,\rm{rad}/\mu$s.}
\label{fig:Arg_deltavarphi}
\end{figure}

Figure~\ref{fig:Arg_deltavarphi} shows the dependence of the phases ${\rm Arg}(j)$ ($j=\alpha, \beta, \zeta, \delta$) and $\delta\varphi\equiv {\rm Arg}(\alpha) + {\rm Arg}(\delta) - {\rm Arg}(\beta) - {\rm Arg}(\zeta)$ on the qubit coupling strength $J$ at a specific time $t=5.325$ $\mu$s, where the populations of the basis states are approximately equal. We note that $\delta\varphi$ reaches $-\pi$ at $J=10^{-3}$ rad/$\mu$s, leading to the maximally entangled state $|\tilde{\psi}\rangle = e^{i\frac{\pi}{4}}( |ff\rangle + |fe\rangle + |ef\rangle -  |ee\rangle)/2$.

\begin{figure}[htp]
\centering
\includegraphics[width=0.68\columnwidth]{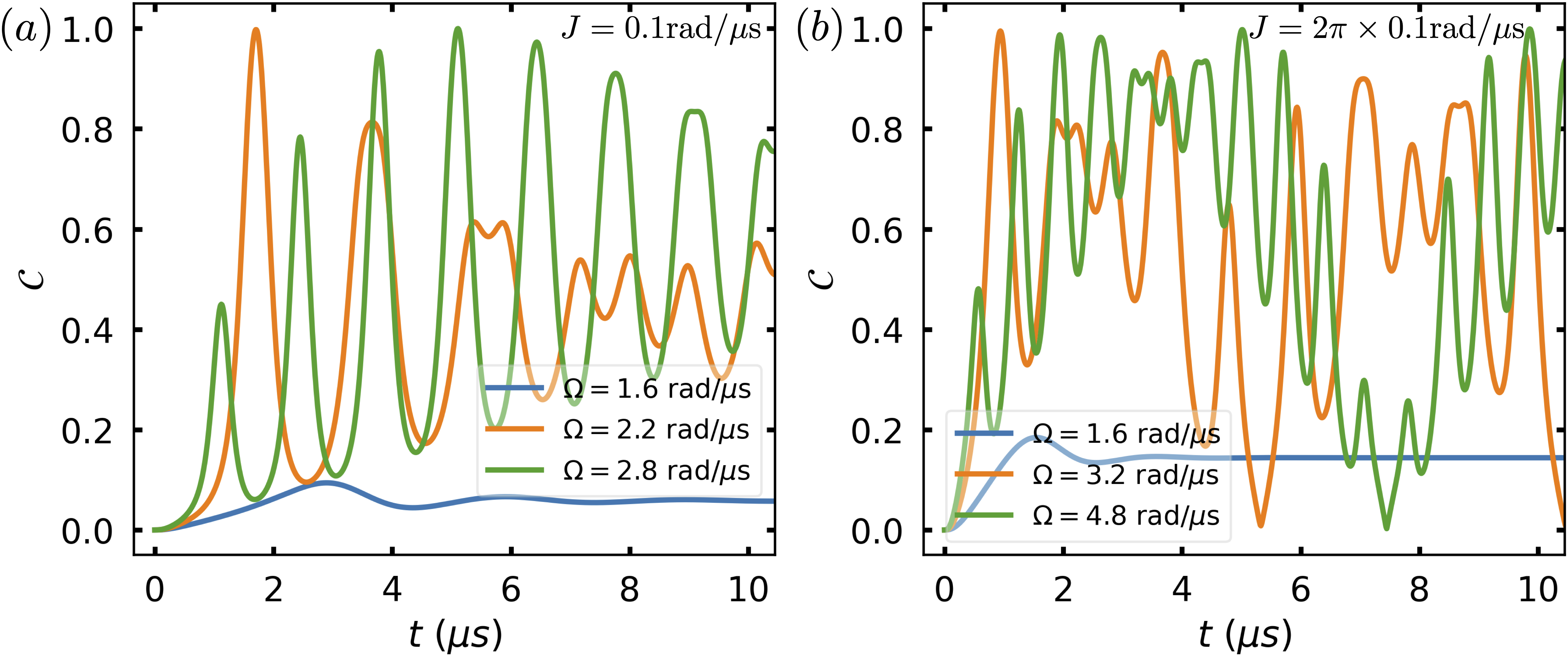} 
\caption{{\bf Entanglement generation at strong qubit coupling.} Concurrence evolution at the $J$ values much larger than that used in the main text: (a) $J=0.1$ rad/$\mu$s and (b) $J=2\pi\times 0.1$ rad/$\mu$s. For each case, we provide the results at three different drive amplitudes $\Omega$. By properly choosing the drive, we can still observe maximal concurrence for these relatively larger $J$ values. The initial state is $|\psi(0)\rangle=|ff\rangle$, and $\gamma=6$ $\mu{\rm s}^{-1}$.
} 
\label{fig:FigS_large_J}
\end{figure}

Figures~\ref{fig:FigS_large_J}(a) and \ref{fig:FigS_large_J}(b) show the concurrence evolution for $J=0.1$ rad/$\mu$s and $J=2\pi\times 0.1$ rad/$\mu$s, respectively. These coupling strengths are much larger than $J = 10^{-3}$ rad$/\mu$s considered in the main text, but we can still obtain a maximally entangled state 
by tuning the drive amplitude to $\Omega=2.2$ rad/$\mu$s [Fig.~\ref{fig:FigS_large_J}(a)] and $\Omega=3.2$ rad/$\mu$s [Fig.~\ref{fig:FigS_large_J}(b)], respectively. 
The time needed to reach maximal concurrence is now comparable to, yet still shorter than that of two coupled Hermitian qubits at the same coupling strength ($\sim 1/J$). 

\section{Non-Hermitian Perturbation theory}


The total Hamiltonian [i.e., Eq.~(2) in the main text], assuming that each qubit is under resonant drive, is given by
\begin{eqnarray}
H &=& H_0 + H_{\rm int} 
\label{eq:total_Hamil}
\end{eqnarray}
with 
\begin{eqnarray}
H_0  &=& \sum_{j=1,2} H_{j}
= \sum_{j=1,2}  -\frac{i\gamma}{2} |e\rangle_j\langle e| + \Omega(|f\rangle_j\langle e| + |e\rangle_j\langle f|) ,  \label{eq:H_0} \\
%
%
H_{\rm int} &=& J(|ef\rangle\langle fe| + |fe\rangle\langle ef|) ,
\label{eq:H_int}
\end{eqnarray}
where $H_0$ is the unperturbed part and $H_{\rm int}$ the interaction Hamiltonian, with a small parameter $J$ representing the perturbation. 
Here we have assumed that the two qubits have the same applied drive and dissipation rate, i.e., $\Omega_1=\Omega_2=\Omega$ and $\gamma_1=\gamma_2=\gamma$.

\subsection{Single non-Hermitian qubit}

We start by reviewing the construction of a single non-Hermitian qubit and explain how to build a dual Hilbert space ${\cal H^{*}}$ for the non-Hermitian qubit in the Hilbert space ${\cal H}$, in order to establish a biorthogonal complete basis. The results can help construct the eigenvalues and eigenstates of the unperturbed Hamiltonian $H_0$ for two non-Hermitian qubits.

The Hamiltonian for each non-Hermitian qubit $H_{j}$ in Eq.~(\ref{eq:H_0}) reads
\begin{eqnarray}
H_{j}=-\frac{i\gamma}{2} |e\rangle_j\langle e| + \Omega(|f\rangle_j\langle e| + |e\rangle_j\langle f|) ,
\label{eq:Hamil_singleQ_Hilbert}
\end{eqnarray}
with eigenvalues 
\begin{eqnarray}
\lambda_{j,\mp} &=& \frac{1}{4} (-i\gamma \mp \sqrt{16\Omega^2 - \gamma^2}) ,
%
\label{eq:eval_indivQ_Hilbert}
\end{eqnarray}
and eigenvalue differences 
$
\delta\lambda_j=\lambda_{j,+}-\lambda_{j,-}=\frac{1}{2}\sqrt{16\Omega^2-\gamma^2}.
$ 
Note that when $\eta \equiv \sqrt{16\Omega^2 -\gamma^2} = 0$, the two eigenvalues $\lambda_{j,\pm}$ become equal.
The eigenstates in this single-qubit Hilbert space ${\cal H}$ are given by
\begin{eqnarray}
|\psi_{j,-} \rangle &=&
\frac{1}{4\sqrt{2}\Omega}
\begin{pmatrix}
i\gamma -\sqrt{16\Omega^2-\gamma^2} \\
4 \Omega
\end{pmatrix} ,
|\psi_{j,+} \rangle =
\frac{1}{4\sqrt{2}\Omega}
\begin{pmatrix}
i\gamma +\sqrt{16\Omega^2-\gamma^2} \\
4 \Omega
\end{pmatrix} .
\label{eq:singleQ_states}
\end{eqnarray}
It is easy to check that the states are normalized, i.e., $\langle \psi_{j,-}| \psi_{j,-}\rangle = \langle \psi_{j,+}| \psi_{j,+}\rangle = 1$, and that they are not  orthogonal, namely,
\begin{eqnarray}
\langle \psi_{j,-}| \psi_{j,+}\rangle \neq 0 .
\label{eq:singleQ_Hilbert_orthoNo}
\end{eqnarray}
This non-orthogonality derives from the non-Hermiticity of  $\cal H$, i.e., from the presence of the terms in $i\gamma$. 
To address this important issue resulting from the non-Hermiticity, we extend the state basis to include states in the dual Hilbert space ${\cal H}^{*}$. 
The Hamiltonian in the dual space ${\cal H^{*}}$ under consideration here is just the Hermitian conjugate of $H_{j}$ in Eq.~(\ref{eq:Hamil_singleQ_Hilbert}), namely
\begin{eqnarray}
\bar{H}_{j} \equiv H^{\dagger}_{j}=\frac{i\gamma}{2} |e\rangle_j\langle e| + \Omega(|f\rangle_j\langle e| + |e\rangle_j\langle f|) .
\end{eqnarray}
$\bar{H}_j$ has eigenvalues 
\begin{eqnarray}
\bar{\lambda}_{j,\mp} &=& \frac{1}{4} (i\gamma \mp\sqrt{16\Omega^2 - \gamma^2}) ,
%
%
\end{eqnarray}
which are complex conjugates of the eigenvalues $\lambda_{j,\mp}$ of the states in the Hilbert space ${\cal H}$, and have eigenvalue differences 
$\delta\bar{\lambda}_j=\bar{\lambda}_{j,+} -\bar{\lambda}_{j,-}=\frac{1}{2}\sqrt{16\Omega^2-\gamma^2} (=\delta\lambda_j)$. 
The eigenstates of $\bar{H}_{j}$ in this dual space are then obtained as
\begin{eqnarray}
|\bar{\psi}_{j,-} \rangle &=&
\frac{1}{4\sqrt{2}\Omega}
\begin{pmatrix}
- i\gamma -\sqrt{16\Omega^2-\gamma^2}) \\
4 \Omega
\end{pmatrix} ,
|\bar{\psi}_{j,+} \rangle =
\frac{1}{4\sqrt{2}\Omega}
\begin{pmatrix}
-i\gamma +\sqrt{16\Omega^2-\gamma^2} \\
4 \Omega
\end{pmatrix} ,
\label{eq:singleQ_states_dual}
\end{eqnarray}
which similarly satisfy $\langle \bar{\psi}_{j,-}| \bar{\psi}_{j,-}\rangle = \langle \bar{\psi}_{j,+}| \bar{\psi}_{j,+}\rangle = 1$ and 
\begin{eqnarray}
\langle \bar{\psi}_{j,-}| \bar{\psi}_{j,+}\rangle \neq 0 .
\label{eq:singleQ_dual_orthoNo}
\end{eqnarray}
The states $|\psi_{j,+}\rangle, |\psi_{j,-}\rangle$ and $\langle\bar{\psi}_{j,+}|, \langle\bar{\psi}_{j,-}|$ are sometimes referred to as the right and left eigenvectors of $H_j$, respectively~\cite{AshidaGongUeda20AdvPhys}.

Although the eigenstates of 
$H_{j}$ and $\bar{H}_{j}$ are not orthogonal in their respective Hilbert and dual Hilbert spaces [Eqs. (\ref{eq:singleQ_Hilbert_orthoNo}) and (\ref{eq:singleQ_dual_orthoNo}), respectively], we can nevertheless construct a biorthogonal basis that is composed of complementary 
pairs of eigenstates in 
$\cal H$ and $\cal H^{*}$ which are orthogonal, specifically,
\begin{eqnarray}
\langle \bar{\psi}_{j,-}| \psi_{j,+}\rangle = \langle \bar{\psi}_{j,+}| \psi_{j,-}\rangle =0 . 
\end{eqnarray}
%
We can then obtain the following normalized biorthogonal eigenstates in the Hilbert space, 
$|\tilde{\psi}_{j,+}\rangle=|\psi_{j,+}\rangle/\sqrt{\langle \bar{\psi}_{j,+}|\psi_{j,+}\rangle}$ and $|\tilde{\psi}_{j,-}\rangle=|\psi_{j,-}\rangle/\sqrt{\langle \bar{\psi}_{j,-}|\psi_{j,-}\rangle}$, where we have used the 
normalization for biorthogonal states. 

This means that the normalized biorthogonal states  cannot be used to reveal the order of the EP, since the overlap factors in the denominators go to zero and the eigenvectors are divergent. Instead, the order of the EP must be confirmed by direct 
diagonalization of the Hamiltonian and analysis of its eigenvectors. 
A related divergence is known in classical systems for the Petermann factor~\cite{Petermann79} which measures the self-overlap of left and right eigenvectors~\cite{Berry03,Lee08pra,Zheng10pra}.  

In the next subsection we extend this biorthogonal basis of single non-Hermitian qubit to the case of two coupled qubits, in order to develop the non-Hermitian perturbation theory for the latter. 

\subsection{Eigenvalues and eigenstates of the unperturbed Hamiltonian $H_0$}

\subsubsection{${\cal PT}$-symmetry preserving phase, $\eta > 0$ \label{app:unbrokenPhase}}

Based on the results of the single non-Hermitian qubit above, we now construct the eigenstates and eigenvalues of the Hamiltonian $H_0$ and its Hermitian conjugate $\bar{H}_0(\equiv H_0^{\dagger})$, 
which describe two non-Hermitian qubits with Hamiltonians $H_j, j=1,2$ in the absence of any qubit coupling. 
Since we are interested in the ${\cal PT}$-symmetry preserving phase ($\Omega>\Omega_{\rm EP}=\gamma/4$) where our main findings are observed, for the sake of simplicity in the following we shall assume 
$\eta>0$, where $\eta$ has been defined above. 

The first two eigenvalues of $H_0$ in Eq. (\ref{eq:H_0}) are given by 
\begin{eqnarray}
\lambda_{--} &=& 2\lambda_{j,-} = -\frac{1}{2} (i\gamma + \eta) , \\
\lambda_{++} &=& 2\lambda_{j,+} = -\frac{1}{2} (i\gamma - \eta) ,
\end{eqnarray}
and the corresponding 
eigenstates in the two-qubit Hilbert space are the product states of the eigenstates of the two qubits: 
\begin{eqnarray}
|\psi_{--}\rangle &=& |\psi_{1,-}\rangle |\psi_{2,-}\rangle = \frac{1}{32\Omega^2}
\begin{pmatrix}
-(\gamma +i\eta)^2 \\
4\Omega(i\gamma -\eta) \\
4\Omega(i\gamma -\eta) \\
16\Omega^2
\end{pmatrix} , 
|\psi_{++}\rangle = |\psi_{1,+}\rangle |\psi_{2,+}\rangle = \frac{1}{32\Omega^2}
\begin{pmatrix}
(i\gamma +\eta)^2 \\
4\Omega(i\gamma +\eta) \\
4\Omega(i\gamma +\eta) \\
16\Omega^2
\end{pmatrix} .
\label{eq:2qubit_states_A}
\end{eqnarray}
With regard to the dual space, the eigenvalues of $\bar{H}_0$ become
\begin{eqnarray}
\bar{\lambda}_{--} &=& 2\bar{\lambda}_{j,-} = \frac{1}{2} (i\gamma - \eta) , \\
\bar{\lambda}_{++} &=& 2\bar{\lambda}_{j,+} = \frac{1}{2} (i\gamma + \eta) ,
\end{eqnarray}
and the associated 
eigenstates are
\begin{eqnarray}
|\bar{\psi}_{--}\rangle &=& |\bar{\psi}_{1,-}\rangle \bar{|\psi}_{2,-}\rangle = \frac{1}{32\Omega^2}
\begin{pmatrix}
-(\gamma -i\eta)^2 \\
-4\Omega(i\gamma +\eta) \\
-4\Omega(i\gamma +\eta) \\
16\Omega^2
\end{pmatrix} , 
|\bar{\psi}_{++}\rangle = |\bar{\psi}_{1,+}\rangle |\bar{\psi}_{2,+}\rangle = \frac{1}{32\Omega^2}
\begin{pmatrix}
-(\gamma +i\eta)^2 \\
4\Omega (\eta-i\gamma ) \\
4\Omega (\eta-i\gamma) \\
16\Omega^2
\end{pmatrix} .
\end{eqnarray}
These four eigenstates 
can now be normalized as described above using biorthogonality as
\begin{eqnarray}
|\tilde{\psi}_{--}\rangle  &=& \frac{|\psi_{--}\rangle}{\sqrt{\langle \bar{\psi}_{--}| \psi_{--}\rangle}}
= \frac{1}{2\eta}
\begin{pmatrix}
-i\gamma+\eta \\
- 4\Omega \\
- 4\Omega \\
i\gamma+\eta
\end{pmatrix} ,
\langle \tilde{\bar{\psi}}_{--}| = \frac{\langle \bar{\psi}_{--}|}{\sqrt{\langle \bar{\psi}_{--}| \psi_{--}\rangle}}
= \frac{1}{2\eta}
\begin{pmatrix}
-i\gamma+\eta, 
- 4\Omega , 
- 4\Omega , 
i\gamma+\eta 
\end{pmatrix} ,
\label{eq:2qubit_states_N--}
\end{eqnarray}
and
\begin{eqnarray}
|\tilde{\psi}_{++}\rangle  &=& \frac{|\psi_{++}\rangle}{\sqrt{\langle \bar{\psi}_{++}| \psi_{++}\rangle}}
= \frac{1}{2\eta}
\begin{pmatrix}
i\gamma+\eta \\
4\Omega \\
4\Omega \\
-i\gamma+\eta
\end{pmatrix} ,
\langle \tilde{\bar{\psi}}_{++}| = \frac{\langle \bar{\psi}_{++}|}{\sqrt{\langle \bar{\psi}_{++}| \psi_{++}\rangle}}
= \frac{1}{2\eta} 
\begin{pmatrix}
i\gamma+\eta,
4\Omega ,
4\Omega ,
-i\gamma+\eta 
\end{pmatrix} .
\label{eq:2qubit_states_N++}
\end{eqnarray}
Here, $\langle\bar{\psi}_{--}|$ and $\langle\bar{\psi}_{++}|$ are the Hermitian conjugates of $|\bar{\psi}_{--}\rangle$ and $|\bar{\psi}_{++}\rangle$ in the dual space, respectively. 
%

The other two eigenvalues of $H_0$ are degenerate and given by
\begin{eqnarray}
\lambda_{-+} &=& \lambda_{+-} = \lambda_{j,-} + \lambda_{j,+} = -\frac{i\gamma}{2}.
\label{eq:eval_degne_Hilbert}
\end{eqnarray}
Their associated 
eigenstates in the corresponding degenerate subspace of $H_0$ are obtained as 
\begin{eqnarray}
|\psi_{-+}\rangle &=& |\psi_{1,-}\rangle |\psi_{2,+}\rangle
= \frac{1}{8\Omega}
\begin{pmatrix}
-4\Omega \\
i\gamma-\eta \\
i\gamma+\eta \\
4\Omega
\end{pmatrix} , 
|\psi_{+-}\rangle = |\psi_{1,+}\rangle |\psi_{2,-}\rangle
= \frac{1}{8\Omega}
\begin{pmatrix}
-4\Omega \\
i\gamma+\eta \\
i\gamma-\eta \\
4\Omega
\end{pmatrix} .
\label{eq:2qubit_states_B}
\end{eqnarray}
For the eigenvalues and 
degenerate eigenstates of $\bar{H}_0$ 
in the dual space, we have 
\begin{eqnarray}
\bar{\lambda}_{-+} &=& \bar{\lambda}_{+-} = \bar{\lambda}_{j,-} + \bar{\lambda}_{j,+} = \frac{i\gamma}{2} ,
\label{eq:eval_degne_Dual}
\end{eqnarray}
and  
\begin{eqnarray}
|\bar{\psi}_{-+} \rangle &=& |\bar{\psi}_{1,-}\rangle |\bar{\psi}_{2,+}\rangle
= \frac{1}{8\Omega}
\begin{pmatrix}
-4\Omega \\
-(i\gamma + \eta) \\
-i\gamma + \eta \\
4\Omega 
\end{pmatrix} , 
|\bar{\psi}_{+-} \rangle = |\bar{\psi}_{1,+}\rangle |\bar{\psi}_{2,-}\rangle
= \frac{1}{8\Omega}
\begin{pmatrix}
-4\Omega \\
-i\gamma + \eta \\
-(i\gamma + \eta) \\
4\Omega 
\end{pmatrix}.
\end{eqnarray} 
Similarly, we can now construct the normalized and unperturbed biorthognal eigenstates in the degenerate subspace 
as
\begin{eqnarray}
|\tilde{\psi}_{-+} \rangle &=& \frac{|\psi_{-+} \rangle}{\sqrt{\langle \bar{\psi}_{-+} |\psi_{-+} \rangle}}
= \frac{1}{2\eta}
\begin{pmatrix}
- 4\Omega \\
i\gamma - \eta \\
i\gamma + \eta \\
4\Omega 
\end{pmatrix} ,
\langle \tilde{\bar{\psi}}_{-+} | = \frac{\langle \bar{\psi}_{-+} |}{\sqrt{\langle \bar{\psi}_{-+} |\psi_{-+} \rangle}}
= \frac{1}{2\eta}
\begin{pmatrix}
-4\Omega,
i\gamma - \eta , 
i\gamma + \eta, 
4\Omega 
\end{pmatrix} ,
\label{eq:2qubit_states_N-+}
\end{eqnarray}
and
\begin{eqnarray}
|\tilde{\psi}_{+-} \rangle &=& \frac{|\psi_{+-} \rangle}{\sqrt{\langle \bar{\psi}_{+-} |\psi_{+-} \rangle}}
= \frac{1}{2\eta}
\begin{pmatrix}
-4\Omega \\
i\gamma + \eta \\
i\gamma - \eta \\
4\Omega
\end{pmatrix} ,
\langle \tilde{\bar{\psi}}_{+-} | = \frac{\langle \bar{\psi}_{+-} | }{\sqrt{\langle \bar{\psi}_{+-} |\psi_{+-} \rangle}}
= \frac{1}{2\eta}
\begin{pmatrix}
-4\Omega,
i\gamma + \eta ,
i\gamma - \eta,
4\Omega
\end{pmatrix} .
\label{eq:2qubit_states_N+-}
\end{eqnarray}
%

In summary, we have constructed a biorthogonal basis for the unperturbed Hamiltonian $H_0$ that is composed of normalized eigenvectors $\{|\tilde{\psi}_{--}\rangle, |\tilde{\psi}_{++}\rangle, |\tilde{\psi}_{-+}\rangle, |\tilde{\psi}_{+-}\rangle\}$ in the Hilbert space and normalized eigenvectors $\{\langle \tilde{\bar{\psi}}_{--}|, \langle \tilde{\bar{\psi}}_{++}|, \langle \tilde{\bar{\psi}}_{-+}|, \langle \tilde{\bar{\psi}}_{+-}|\}$ (or $\{|\tilde{\bar{\psi}}_{--}\rangle, |\tilde{\bar{\psi}}_{++}\rangle, |\tilde{\bar{\psi}}_{-+}\rangle, |\tilde{\bar{\psi}}_{+-}\rangle\}$) in the dual Hilbert space. 
It is straightforward to verify these states are not only normalized, i.e.,
\begin{eqnarray}
\langle \tilde{\bar{\psi}}_{--} |\tilde{\psi}_{--}\rangle =\langle \tilde{\bar{\psi}}_{++} |\tilde{\psi}_{++}\rangle =\langle \tilde{\bar{\psi}}_{-+} |\tilde{\psi}_{-+}\rangle =\langle \tilde{\bar{\psi}}_{+-} |\tilde{\psi}_{+-}\rangle= 1 ,
\end{eqnarray}
but also orthogonal, with
\begin{eqnarray}
\langle \tilde{\bar{\psi}}_{--} |\tilde{\psi}_{++}\rangle 
&=& \langle \tilde{\bar{\psi}}_{-+} |\tilde{\psi}_{++}\rangle 
= \langle \tilde{\bar{\psi}}_{+-} |\tilde{\psi}_{++}\rangle 
= 0 , \\
\langle \tilde{\bar{\psi}}_{++} |\tilde{\psi}_{--}\rangle 
&=& \langle \tilde{\bar{\psi}}_{-+} |\tilde{\psi}_{--}\rangle 
= \langle \tilde{\bar{\psi}}_{+-} |\tilde{\psi}_{--}\rangle 
=0 ,\\
\langle \tilde{\bar{\psi}}_{++} |\tilde{\psi}_{-+}\rangle 
&=& \langle \tilde{\bar{\psi}}_{--} |\tilde{\psi}_{-+}\rangle 
= \langle \tilde{\bar{\psi}}_{+-} |\tilde{\psi}_{-+}\rangle 
=0 ,\\
\langle \tilde{\bar{\psi}}_{++} |\tilde{\psi}_{+-}\rangle 
&=& \langle \tilde{\bar{\psi}}_{--} |\tilde{\psi}_{+-}\rangle 
= \langle \tilde{\bar{\psi}}_{-+} |\tilde{\psi}_{+-}\rangle 
=0 .
\end{eqnarray}

We now consider the implications for characterization of the EP in this two-qubit system.  First, we note that all four eigenvalues of
$H_0$, namely
$\lambda_{++}, \lambda_{--}, \lambda_{+-}, \lambda_{-+}$, are equal to $-i\gamma/2$ when $\eta=0$, suggesting that this is a candidate location for an EP. 
Note that as for the single qubit case in Section II.A, these normalized biorthogonal eigenvectors in the Hilbert space [i.e., $|\tilde{\psi}_{--}\rangle$, $|\tilde{\psi}_{++}\rangle$, $|\tilde{\psi}_{-+}\rangle$, and $|\tilde{\psi}_{+-}\rangle$ presented in Eqs.~(\ref{eq:2qubit_states_N--}), (\ref{eq:2qubit_states_N++}), (\ref{eq:2qubit_states_N-+}), and  (\ref{eq:2qubit_states_N+-}), respectively] will diverge at the EP 
and thus cannot be used to determine the order of this. 
In this two-qubit case we have specified above also the conventionally-normalized 
states $|\psi_{--}\rangle$, $|\psi_{++}\rangle$, $|\psi_{-+}\rangle$, and $|\psi_{+-}\rangle$ in Eqs.~(\ref{eq:2qubit_states_A}) and (\ref{eq:2qubit_states_B}). There are also all identical when $\eta=0$: taken together with the equal eigenvalues, this indicates a fourth-order EP. 
It is essential that the equivalence of eigenvectors is between normalized eigenvectors of the Hamiltonian $H_0$, as was done for the single qubit in Section II.A above. In general, these can be readily obtained by analytical or numerical diagonalization of $H_0$.  Once conventionally-normalized eigenvectors of $H_0$ at a candidate EP are obtained, whether numerically or analytically, one can then evaluate the overlap of eigenstates. These are shown for the two-qubit Hamiltonian in Fig.~\ref{fig:Overlap_eigenstates}(a) of Sec.~\ref{app:figures} and unambiguously confirms that in this case the EP is fourth order since all four of the normalized eigenvectors are identical. 

\subsubsection{${\cal PT}$-symmetry broken phase, $\eta < 0$ \label{app:brokenPhase}}

While we mainly focus on the ${\cal PT}$-symmetry preserving phase throughout our work, the illustration of the EP order calculated from the overlap of eigenvectors $|\psi_i\rangle$ shown in Fig.~\ref{fig:Overlap_eigenstates}(a) contains the broken-phase regime. To this purpose, we present the conventionally normalized eigenvectors $|\psi_i\rangle$ in the unbroken phase, namely 
\begin{eqnarray}
|\psi_{--}\rangle &=& \frac{1}{32\Omega^2}
\begin{pmatrix}
(i\gamma +\eta)^2 \\
4\Omega(i\gamma +\eta) \\
4\Omega(i\gamma +\eta) \\
16\Omega^2
\end{pmatrix} ,
|\psi_{++}\rangle = \frac{1}{32\Omega^2}
\begin{pmatrix}
-(\gamma +i\eta)^2 \\
4\Omega(i\gamma -\eta) \\
4\Omega(i\gamma -\eta) \\
16\Omega^2
\end{pmatrix} ,
\end{eqnarray}
and 
\begin{eqnarray}
|\psi_{-+}\rangle &=& \frac{1}{8\Omega}
\begin{pmatrix}
-4\Omega \\
i\gamma+\eta \\
i\gamma-\eta \\
4\Omega)^T
\end{pmatrix} ,  
|\psi_{+-}\rangle = \frac{1}{8\Omega}
\begin{pmatrix}
-4\Omega \\
i\gamma-\eta \\
i\gamma+\eta \\
4\Omega
\end{pmatrix} , 
\end{eqnarray}
with corresponding eigenvalues given by $\lambda_{--} = \frac{1}{2} (\eta - i\gamma)$, $\lambda_{++} = -\frac{1}{2} (\eta + i\gamma)$, and $\lambda_{-+} = \lambda_{+-} = -\frac{i}{2} \gamma$, respectively. 

\subsection{First-order degenerate and non-degenerate perturbation theory}

When a non-Hermitian degeneracy exists in the  Hilbert space under consideration or its dual space [see Eqs.~(\ref{eq:eval_degne_Hilbert}) and (\ref{eq:eval_degne_Dual})], a direct generalization of the conventional perturbation theory to the non-Hermitian systems does not work. Following the idea of degenerate perturbation theory in standard Hermitian quantum mechanics, we shall first lift the degeneracy by using the Hermitian perturbation $H_{\rm int}$ [see Eq. (\ref{eq:H_int})], i.e., the diagonalization of the interaction Hamiltonian in the degenerate subspace. 

The perturbation matrix in the subspace spanned by $\{|\tilde{\psi}_{+-}\rangle, |\tilde{\psi}_{-+}\rangle \}$ and their adjoint states $\{ \langle \tilde{\bar{\psi}}_{+-}|, \langle \tilde{\bar{\psi}}_{-+}|\}$ can be obtained as
\begin{eqnarray}
[H_{\rm int}] &\equiv& 
\begin{pmatrix}
\langle \tilde{\bar{\psi}}_{+-}| H_{\rm int} |\tilde{\psi}_{+-}\rangle & \langle \tilde{\bar{\psi}}_{-+}| H_{\rm int} |\tilde{\psi}_{+-}\rangle \\
\langle \tilde{\bar{\psi}}_{-+}| H_{\rm int} |\tilde{\psi}_{+-}\rangle & \langle \tilde{\bar{\psi}}_{-+}| H_{\rm int} |\tilde{\psi}_{-+}\rangle \\
\end{pmatrix}
=
\begin{pmatrix}
-\frac{8J\Omega^2}{\eta^2} & J-\frac{8J\Omega^2}{\eta^2} \\
J-\frac{8J\Omega^2}{\eta^2} & -\frac{8J\Omega^2}{\eta^2}
\end{pmatrix} .
\label{eq:submatrix}
\end{eqnarray}
We note that since $H_{\rm int}$ is Hermitian, this submatrix is real  and has eigenvalues $-J$ and $J-\frac{16J\Omega^2}{\eta^2}$ and eigenstates $(|\psi_{-+}\rangle \mp |\psi_{+-})/\sqrt{2}$.
This suggests that we can choose a new basis for the degenerate subspace, namely
\begin{eqnarray}
|\tilde{\psi}_{1} \rangle&=&\frac{1}{\sqrt{2}} (|\tilde{\psi}_{-+}\rangle - |\tilde{\psi}_{+-})
= \frac{1}{\sqrt{2}}
\begin{pmatrix}
0 \\
- 1\\
 1 \\
0
\end{pmatrix} , \nonumber
\\
|\tilde{\psi}_{2} \rangle&=&\frac{1}{\sqrt{2}} (|\tilde{\psi}_{-+}\rangle + |\tilde{\psi}_{+-})
= 
\frac{1}{\sqrt{2}\eta}
\begin{pmatrix}
-4\Omega \\
i\gamma \\
i\gamma \\
4\Omega
\end{pmatrix} , \\
|\tilde{\bar{\psi}}_1\rangle &=& \frac{1}{\sqrt{2}} (|\tilde{\bar{\psi}}_{-+}\rangle -|\tilde{\bar{\psi}}_{+-}\rangle)
= \frac{1}{\sqrt{2}}
\begin{pmatrix}
0 \\
-1 \\
1 \\
0
\end{pmatrix} , \nonumber
\\
|\tilde{\bar{\psi}}_2\rangle &=& \frac{1}{\sqrt{2}} (|\tilde{\bar{\psi}}_{-+}\rangle + |\tilde{\bar{\psi}}_{+-}\rangle)
=
\frac{1}{\sqrt{2}\eta}
\begin{pmatrix}
-4\Omega \\
-i\gamma \\
-i\gamma  \\
4\Omega
\end{pmatrix} ,
\end{eqnarray}
in which the interaction Hamiltonian $H_{\rm int}$ is now diagonal (i.e., with off-diagonal elements being zeros). 
Note that the state $|\tilde{\psi}_{1} \rangle$ is equal to its dual state $|\tilde{\bar{\psi}}_1\rangle$. 
The corresponding eigenvalues are given by
\begin{eqnarray}
\lambda_1 &=& \lambda_{-+}, \lambda_2=\lambda_{+-}, \\
\bar{\lambda}_1 &=& \bar{\lambda}_{-+}, \bar{\lambda}_2=\bar{\lambda}_{+-} .
\end{eqnarray}

We are now ready to apply first-order non-degenerate perturbation theory to our system of two weakly coupled non-Hermitian qubits.
For the degenerate subspace, the first-order corrections to the eigenvalues are given by the eigenvalues of the submatrix in Eq.~(\ref{eq:submatrix}). Therefore, the perturbed eigenvalues are
\begin{eqnarray}
\Lambda_{1} &=& \lambda_{1} -J = -J - \frac{i\gamma}{2}, \label{eq:Lambda_1}\\
\bar{\Lambda}_1 &=& \bar{\lambda}_{1} -J = -J+\frac{i\gamma}{2} , \\
\Lambda_{2} &=& \lambda_{2} + J-\frac{16J\Omega^2}{\eta^2} = -\frac{i\gamma}{2} - \frac{J\gamma^2}{\eta^2} , \label{eq:Lambda_2}\\
\bar{\Lambda}_2 &=&  \bar{\lambda}_{2} + J - \frac{16J\Omega^2}{\eta^2} = \frac{i\gamma}{2} - \frac{J\gamma^2}{\eta^2} . 
\end{eqnarray}
For the corresponding perturbed eigenstates, given that the interaction Hamiltonian is already diagonal in the basis $\{|\tilde{\psi}_1\rangle, |\tilde{\psi}_2\rangle\}$ of the Hilbert space and $\{\langle\tilde{\bar{\psi}}_1|, \langle\tilde{\bar{\psi}}_2|\}$ of the dual space, we therefore only need to include the first-order correction from the unperturbed eigenstates of the non-degenerate subspace. The perturbed eigenstates are then given by
\begin{eqnarray}
|\Psi_{1} \rangle&=&
|\tilde{\psi}_1\rangle 
+ \frac{\langle \tilde{\bar{\psi}}_{--}| H_{\rm int} |\tilde{\psi}_{1}\rangle}{\lambda_{1}-\lambda_{--}} |\tilde{\psi}_{--}\rangle
+ \frac{\langle \tilde{\bar{\psi}}_{++}| H_{\rm int} |\tilde{\psi}_{1}\rangle}{\lambda_{1}-\lambda_{++}} |\tilde{\psi}_{++}\rangle
= \frac{1}{\sqrt{2}}
\begin{pmatrix}
0 \\
- 1\\
 1 \\
0
\end{pmatrix} , \label{eq:Psi_1}
\end{eqnarray}
\begin{eqnarray}
|\bar{\Psi}_1\rangle &=& 
|\tilde{\bar{\psi}}_1\rangle 
+ \frac{ \langle \tilde{\psi}_{--}| H_{\rm int} |\tilde{\bar{\psi}}_{1}\rangle}{\bar{\lambda}_{1}-\bar{\lambda}_{--}} |\tilde{\bar{\psi}}_{--}\rangle
+ \frac{ \langle \tilde{\psi}_{++}| H_{\rm int} |\tilde{\bar{\psi}}_{1}\rangle}{\bar{\lambda}_{1}-\bar{\lambda}_{++}} |\tilde{\bar{\psi}}_{++}\rangle
= \frac{1}{\sqrt{2}}
\begin{pmatrix}
0 \\
-1 \\
1 \\
0
\end{pmatrix} , \label{eq:barPsi_1}
\end{eqnarray}
\begin{eqnarray}
|\Psi_{2} \rangle&=&
|\tilde{\psi_2}\rangle 
+ \frac{\langle \tilde{\bar{\psi}}_{--}| H_{\rm int} |\tilde{\psi}_{2}\rangle}{\lambda_{2}-\lambda_{--}} |\tilde{\psi}_{--}\rangle
+ \frac{\langle \tilde{\bar{\psi}}_{++}| H_{\rm int} |\tilde{\psi}_{2}\rangle}{\lambda_{2}-\lambda_{++}} |\tilde{\psi}_{++}\rangle
= 
\frac{1}{\sqrt{2}\eta^3}
\begin{pmatrix}
-4\Omega(\eta^2+2iJ\gamma) \\
i\gamma\eta^2 \\
i\gamma\eta^2 \\
4\Omega(\eta^2-2iJ\gamma) 
\end{pmatrix} , \label{eq:Psi_2}
\end{eqnarray}
\begin{eqnarray}
|\bar{\Psi}_2\rangle &=& 
|\tilde{\bar{\psi}}_2\rangle 
+ \frac{ \langle \tilde{\psi}_{--}| H_{\rm int} |\tilde{\bar{\psi}}_{2}\rangle}{\bar{\lambda}_{2}-\bar{\lambda}_{--}} |\tilde{\bar{\psi}}_{--}\rangle
+ \frac{ \langle \tilde{\psi}_{++}| H_{\rm int} |\tilde{\bar{\psi}}_{2}\rangle}{\bar{\lambda}_{2}-\bar{\lambda}_{++}} |\tilde{\bar{\psi}}_{++}\rangle
=
\frac{1}{\sqrt{2}\eta^3}
\begin{pmatrix}
-4\Omega(\eta^2-2iJ\gamma) \\
-i\gamma\eta^2 \\
-i\gamma\eta^2 \\
4\Omega(\eta^2+2iJ\gamma) 
\end{pmatrix} . \label{eq:barPsi_2}
\end{eqnarray}
It should be noted that $|\Psi_{1} \rangle$ (or $|\bar{\Psi}_{1} \rangle$) is the same as $|\tilde{\psi}_1\rangle$ (or $|\tilde{\bar{\psi}}_1\rangle$), although a different notation is chosen. 
This indicates that the unperturbed state $|\tilde{\psi}_1\rangle$ (or $|\tilde{\bar{\psi}}_1\rangle$) from the destructive quantum interference between $|\tilde{\psi}_{-+}\rangle$ and $|\tilde{\psi}_{+-}\rangle$ (or $|\tilde{\bar{\psi}}_{-+}\rangle$ and $|\tilde{\bar{\psi}}_{+-}\rangle$) is 
decoupled from all other states induced by $H_{\rm int}$.
Apart from this decoupled state, the other two newly constructed states $|\Psi_{2} \rangle$ and $|\bar{\Psi}_{2} \rangle$ include non-zero first-order corrections to the states $\tilde{|\psi_2}\rangle$ and $|\tilde{\bar{\psi}}_2\rangle$. 

For the non-degenerate subspaces, the first-order corrections are given by the expectation values of the interaction Hamiltonian in the unperturbed eigenstates. 
The perturbed eigenvalues are given by
\begin{eqnarray}
\Lambda_{--} &=& \lambda_{--} + \langle \tilde{\bar{\psi}}_{--}| H_{\rm int} |\tilde{\psi}_{--}\rangle
=-i\frac{\gamma}{2} - \frac{\eta}{2} + \frac{8J\Omega^2}{\eta^2} , \label{eq:Lambda_--}\\
\bar{\Lambda}_{--} &=& \bar{\lambda}_{--} + \langle \tilde{\psi}_{--}| H_{\rm int} |\tilde{\bar{\psi}}_{--}\rangle
=i\frac{\gamma}{2} - \frac{\eta}{2} + \frac{8J\Omega^2}{\eta^2} ,\\
\Lambda_{++} &=& \lambda_{++} + \langle \tilde{\bar{\psi}}_{++}| H_{\rm int} |\tilde{\psi}_{++}\rangle
= -\frac{i\gamma}{2} + \frac{\eta}{2} + \frac{8J\Omega^2}{\eta^2} , \label{eq:Lambda_++}\\
\bar{\Lambda}_{++} &=& \bar{\lambda}_{++} + \langle \tilde{\psi}_{++}| H_{\rm int} |\tilde{\bar{\psi}}_{++}\rangle
= \frac{i\gamma}{2} + \frac{\eta}{2} + \frac{8J\Omega^2}{\eta^2} .
\end{eqnarray}
The corresponding perturbed eigenstates are given by 
\begin{eqnarray}
|\Psi_{--}\rangle &=& |\tilde{\psi}_{--}\rangle 
+ \frac{\langle \tilde{\bar{\psi}}_{++}| H_{\rm int} |\tilde{\psi}_{--}\rangle}{\lambda_{--}-\lambda_{++}} |\tilde{\psi}_{++}\rangle
+ \frac{\langle \tilde{\bar{\psi}}_{1}| H_{\rm int} |\tilde{\psi}_{--}\rangle}{\lambda_{--}-\lambda_{1}} |\tilde{\psi}_{1}\rangle
+ \frac{\langle \tilde{\bar{\psi}}_{2}| H_{\rm int} |\tilde{\psi}_{--}\rangle}{\lambda_{--}-\lambda_{2}} |\tilde{\psi}_{2}\rangle \notag\\
&=& \frac{1}{2\eta^4}
\begin{pmatrix}
\eta^3(\eta-i\gamma) +8J\Omega^2(\eta-3i\gamma) \\
-4\Omega(\eta^3-2J\eta^2+24J\Omega^2) \\
-4\Omega(\eta^3-2J\eta^2+24J\Omega^2) \\
\eta^3(\eta+i\gamma) +8J\Omega^2(\eta+3i\gamma) 
\end{pmatrix} , \label{eq:Psi_--}
\end{eqnarray}
\begin{eqnarray}
|\bar{\Psi}_{--}\rangle &=& |\tilde{\bar{\psi}}_{--}\rangle 
+ \frac{\langle \tilde{\psi}_{++}| H_{\rm int} |\tilde{\bar{\psi}}_{--}\rangle}{\bar{\lambda}_{--}-\bar{\lambda}_{++}} |\tilde{\bar{\psi}}_{++}\rangle
+ \frac{\langle \tilde{\psi}_{1}| H_{\rm int} |\tilde{\bar{\psi}}_{--}\rangle}{\bar{\lambda}_{--}-\bar{\lambda}_{1}} |\tilde{\bar{\psi}}_{1}\rangle
+ \frac{\langle \tilde{\psi}_{2}| H_{\rm int} |\tilde{\bar{\psi}}_{--}\rangle}{\bar{\lambda}_{--}-\bar{\lambda}_{2}} |\tilde{\bar{\psi}}_{2}\rangle \notag\\
&=& \frac{1}{2\eta^4}
\begin{pmatrix}
\eta^3(\eta+i\gamma) +8J\Omega^2(\eta+3i\gamma) \\
-4\Omega(\eta^3-2J\eta^2+24J\Omega^2) \\
-4\Omega(\eta^3-2J\eta^2+24J\Omega^2) \\
\eta^3(\eta-i\gamma) +8J\Omega^2(\eta-3i\gamma) 
\end{pmatrix} , \label{eq:barPsi_--}
\end{eqnarray}
\begin{eqnarray}
|\Psi_{++}\rangle &=& |\tilde{\psi}_{++}\rangle 
+ \frac{ \langle \tilde{\bar{\psi}}_{--}| H_{\rm int} |\tilde{\psi}_{++}\rangle}{\lambda_{++}-\lambda_{--}} |\tilde{\psi}_{--}\rangle
+ \frac{ \langle \tilde{\bar{\psi}}_{1}| H_{\rm int} |\tilde{\psi}_{++}\rangle}{\lambda_{++}-\lambda_{1}} |\tilde{\psi}_{1}\rangle
+ \frac{ \langle \tilde{\bar{\psi}}_{2}| H_{\rm int} |\tilde{\psi}_{++}\rangle}{\lambda_{++}-\lambda_{2}} |\tilde{\psi}_{2}\rangle \notag\\
&=& \frac{1}{2\eta^4}
\begin{pmatrix}
\eta^3(\eta+i\gamma) -8J\Omega^2(\eta+3i\gamma) \\
4\Omega(\eta^3+2J\eta^2-24J\Omega^2) \\
4\Omega(\eta^3+2J\eta^2-24J\Omega^2) \\
\eta^3(\eta-i\gamma) -8J\Omega^2(\eta-3i\gamma) 
\end{pmatrix} , \label{eq:Psi_++}
\end{eqnarray}
\begin{eqnarray}
|\bar{\Psi}_{++}\rangle &=& |\tilde{\bar{\psi}}_{++}\rangle 
+ \frac{ \langle \tilde{\psi}_{--}| H_{\rm int} |\tilde{\bar{\psi}}_{++}\rangle}{\bar{\lambda}_{++}-\bar{\lambda}_{--}} |\tilde{\bar{\psi}}_{--}\rangle
+ \frac{ \langle \tilde{\psi}_{1}| H_{\rm int} |\tilde{\bar{\psi}}_{++}\rangle}{\bar{\lambda}_{++}-\bar{\lambda}_{1}} |\tilde{\bar{\psi}}_{1}\rangle
+ \frac{ \langle \tilde{\psi}_{2}| H_{\rm int} |\tilde{\bar{\psi}}_{++}\rangle}{\bar{\lambda}_{++}-\bar{\lambda}_{2}} |\tilde{\bar{\psi}}_{2}\rangle \notag\\
&=& \frac{1}{2\eta^4}
\begin{pmatrix}
\eta^3(\eta-i\gamma) -8J\Omega^2(\eta-3i\gamma) \\
4\Omega(\eta^3+2J\eta^2-24J\Omega^2) \\
4\Omega(\eta^3+2J\eta^2-24J\Omega^2) \\
\eta^3(\eta+i\gamma) -8J\Omega^2(\eta+3i\gamma) 
\end{pmatrix} . \label{eq:barPsi_++}
\end{eqnarray} 
It is clear that these newly constructed states include the first-order corrections from both non-degenerate and degenerate subspaces to the unperturbed states $|\tilde{\psi}_{--}\rangle$ and $|\tilde{\psi}_{++}\rangle$ of the Hilbert space and to the unperturbed states $|\tilde{\bar{\psi}}_{--}\rangle$ and $|\tilde{\bar{\psi}}_{++}\rangle$ of the dual space .

The above perturbation theory 
provides the basis states (i.e., $|\Psi_{++}\rangle$, $|\Psi_{--}\rangle$, $|\Psi_{1}\rangle$, and $|\Psi_{2}\rangle$) in the Hilbert space ${\cal H}$ as well as the corresponding adjoint states (i.e., $\langle \bar{\Psi}_{++}|$, $\langle\bar{\Psi}_{--}|$, $\langle\bar{\Psi}_1|$, and $\langle\bar{\Psi}_2|$) in the dual space ${\cal H}^*$. 
We verify that, up to the first order in the inter-qubit coupling $J$, these basis states are not only normalized, i.e.,
\begin{eqnarray}
\langle \bar{\Psi}_{++} | \Psi_{++}\rangle &=&  \langle \bar{\Psi}_{--} | \Psi_{--}\rangle = \langle \bar{\Psi}_{2} | \Psi_{2}\rangle =1 +{\cal O}(J^2) ,\\
\langle \bar{\Psi}_{1} | \Psi_{1}\rangle &=& 1 ,
\end{eqnarray}
but also orthogonal, namely
\begin{eqnarray}
\langle \bar{\Psi}_{--} | \Psi_{++}\rangle
&=& \langle \bar{\Psi}_{++} | \Psi_{--}\rangle
= \langle \bar{\Psi}_{2} | \Psi_{++}\rangle 
= \langle \bar{\Psi}_{++} | \Psi_{2}\rangle
= \langle \bar{\Psi}_{2} | \Psi_{--}\rangle
= \langle \bar{\Psi}_{--} | \Psi_{2}\rangle
= {\cal O}(J^2) , \\
\langle \bar{\Psi}_{1} | \Psi_{++}\rangle
&=& \langle \bar{\Psi}_{++} | \Psi_{1}\rangle
= \langle \bar{\Psi}_{1} | \Psi_{--}\rangle
= \langle \bar{\Psi}_{--} | \Psi_{1}\rangle
= \langle \bar{\Psi}_{2} | \Psi_{1}\rangle 
= \langle \bar{\Psi}_{1} | \Psi_{2}\rangle 
= 0 .
\end{eqnarray} 
They also compose a complete basis
\begin{eqnarray}
| \Psi_{++}\rangle\langle \bar{\Psi}_{++}| + | \Psi_{--}\rangle\langle \bar{\Psi}_{--} |
+  | \Psi_{1}\rangle \langle \bar{\Psi}_{1}| +  | \Psi_{2}\rangle\langle \bar{\Psi}_{2}| = 1 + {\cal O}(J^2) .
\end{eqnarray}

Below we list the energy separations 
between the perturbed eigenstates in the Hilbert space
\begin{eqnarray}
\delta_{1,--} &\equiv& \Lambda_1-\Lambda_{--} = \frac{1}{2} \Big( \eta - J\frac{3\eta^2+\gamma^2}{\eta^2} \Big) , \label{eq:delta1--} \\
\delta_{2,--} &\equiv& \Lambda_2-\Lambda_{--} = \frac{1}{2} \Big( \eta -J\frac{\eta^2+3\gamma^2}{\eta^2} \Big), \label{eq:delta2--} \\
\delta_{++,1} &\equiv& \Lambda_{++}-\Lambda_1 = \frac{1}{2} \Big( \eta + J\frac{3\eta^2+\gamma^2}{\eta^2} \Big), \label{eq:delta++1} \\
\delta_{++,2} &\equiv& \Lambda_{++}-\Lambda_2 = \frac{1}{2} \Big( \eta + J\frac{\eta^2+3\gamma^2}{\eta^2} \Big) .
\label{eq:delta++2} 
\end{eqnarray}
These are shown in Fig.~\ref{fig:schematic_crossResonantDrive} which provides a schematic illustration of the 
perturbed energy structure of the non-Hermitian two-qubit system.

\subsection{State evolution and entanglement}

Given an initial state $|\psi(0)\rangle$, the time evolution of the state under an evolution operator $U=e^{-iHt}$ 
can be given by
\begin{eqnarray}
|\psi(t)\rangle &=& U|\psi(0)\rangle
= U (| \Psi_{++}\rangle\langle \bar{\Psi}_{++}| + | \Psi_{--}\rangle\langle \bar{\Psi}_{--} |
+  | \Psi_{1}\rangle\langle \bar{\Psi}_{1}| + | \Psi_{2}\rangle\langle \bar{\Psi}_{2}| ) |\psi(0)\rangle \notag\\
&=&
\langle \bar{\Psi}_{++} |\psi(0)\rangle \cdot U | \Psi_{++}\rangle
+ \langle \bar{\Psi}_{--} |\psi(0)\rangle \cdot U | \Psi_{--}\rangle
+ \langle \bar{\Psi}_{1} |\psi(0)\rangle \cdot U | \Psi_{1}\rangle
+ \langle \bar{\Psi}_{2} |\psi(0)\rangle \cdot U | \Psi_{2}\rangle \notag\\
&=&
\langle \bar{\Psi}_{++} |\psi(0)\rangle \cdot e^{-it\Lambda_{++}} | \Psi_{++}\rangle
+ \langle \bar{\Psi}_{--} |\psi(0)\rangle \cdot e^{-it\Lambda_{--}} | \Psi_{--}\rangle \notag\\
&&
+ \langle \bar{\Psi}_{1} |\psi(0)\rangle \cdot e^{-it\Lambda_{1}} | \Psi_{1}\rangle
+ \langle \bar{\Psi}_{2} |\psi(0)\rangle \cdot e^{-it\Lambda_{2}} | \Psi_{2}\rangle.
\label{eq:state}
\end{eqnarray}
For simplicity, we shall write $|\psi(t)\rangle = (\alpha^{\prime}, \beta^{\prime}, \zeta^{\prime}, \delta^{\prime})^T$, which implicitly defines the time-dependent quantities $\alpha^{\prime}, \beta^{\prime}, \zeta^{\prime}, \delta^{\prime}$. The according normalized state is given by
\begin{eqnarray}
|\tilde{\psi}(t)\rangle &=& \frac{|\psi(t)\rangle}{| |\psi(t)\rangle |} . \label{eq:stateNormlized}
\end{eqnarray}


By projecting this two-qubit pure state onto the four maximally entangled Bell states, i.e., $|\psi\rangle=\sum_j c_j|e_j\rangle$ with 
\begin{eqnarray}
|e_1\rangle &=& \frac{1}{\sqrt{2}}(|ff\rangle +|ee\rangle), 
|e_2\rangle = \frac{i}{\sqrt{2}}(|ff\rangle -|ee\rangle), \label{eq:Bell_++--}\\
|e_3\rangle &=& \frac{i}{\sqrt{2}}(|fe\rangle +|ef\rangle), 
|e_4\rangle = \frac{1}{\sqrt{2}}(|fe\rangle -|ef\rangle), \label{eq:Bell_-++-}
\end{eqnarray}
the concurrence as an entanglement measure can then be written as
\begin{eqnarray}
\cal{C} &=& \frac{|\langle \psi^*|\psi \rangle| }{||\psi(t)\rangle|^2}
= \frac{|\sum_j c_j^2| }{||\psi(t)\rangle|^2}
= \frac{2|\alpha^{\prime}\delta^{\prime} -\beta^{\prime}\zeta^{\prime}|}{|\alpha^{\prime}|^2+|\beta^{\prime}|^2+|\zeta^{\prime}|^2+|\delta^{\prime}|^2} ,
\label{eq:C_analytics}
\end{eqnarray}
where $|\psi^*\rangle=\sum_j c_j^*|e_j\rangle$. 
The explicit expression for $\cal{C}$ as well as those for $\alpha^{\prime}, \beta^{\prime}, \zeta^{\prime}, \delta^{\prime}$ is very lengthy and not included here, but simplified versions under a further approximation will be provided below. 

In Fig. \ref{fig:analytics_approx_numerics}, we compare the results calculated from Eq.~(\ref{eq:C_analytics}) with the exact solutions from fully numerical calculations for three different values of $\Omega$, where the initial state is given by $|\psi(0)\rangle=|ff\rangle$. The results show that our perturbative theory assuming the weak qubit coupling up to first order is in good agreement with numerically exact calculations.


\subsection{Approximate solutions}

\begin{figure}
\centering
  \includegraphics[width=.69\columnwidth]{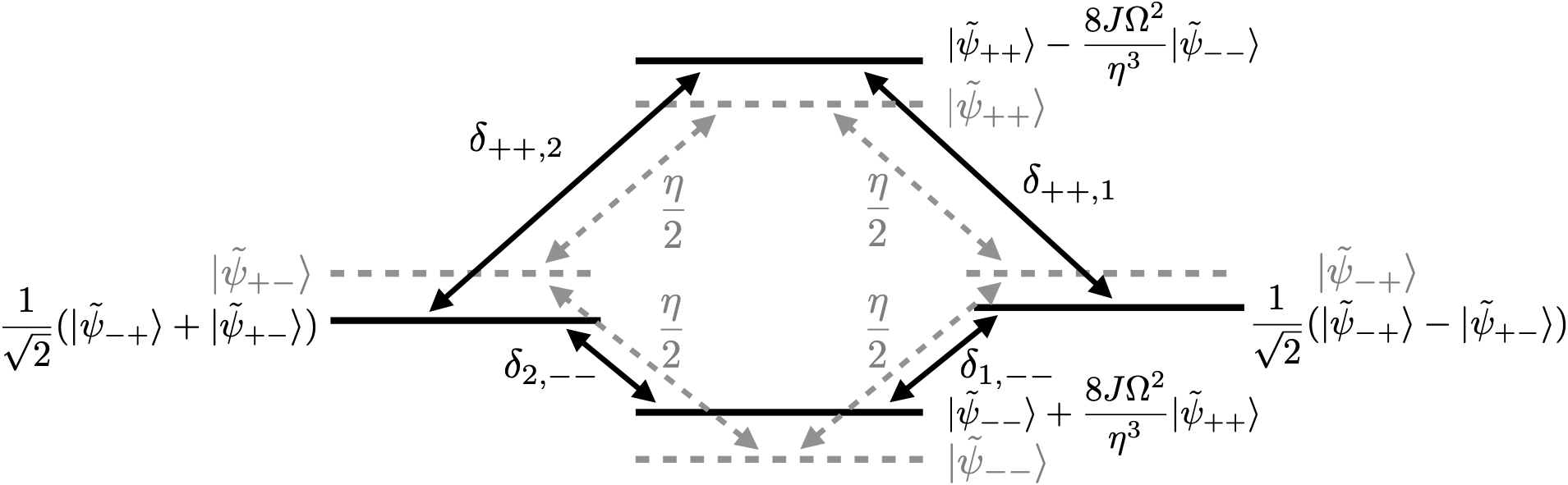}
\caption{
{\bf Approximate energy levels of weakly coupled non-Hermitian qubits.}
The perturbed states in the Hilbert space are approximately given by Eqs. (\ref{eq:Psi_--_Approx}), (\ref{eq:Psi_++_Approx}), (\ref{eq:Psi_1_Approx}), (\ref{eq:Psi_2_Approx}) and the associated transition frequencies, i.e., $\delta_{++,1}$, $\delta_{++,2}$, $\delta_{1,--}$, and $\delta_{2,--}$ follow Eqs. (\ref{eq:delta1--})-(\ref{eq:delta++2}). 
The unperturbed energy levels are presented as grey dashed lines. There is a common energy gap $\frac{\eta}{2}$ between $|\tilde{\psi}_{++ (--)}\rangle$  and $|\tilde{\psi}_{+-}\rangle$ or $|\tilde{\psi}_{-+}\rangle$. 
The perturbation increases or decreases this gap by an amount of $J\frac{\eta^2+3\gamma^2}{\eta^2}$ leading to $\delta_{++,2}$ or $\delta_{2,--}$, respectively, while this shift becomes $J\frac{3\eta^2+\gamma^2}{\eta^2}$ for $\delta_{++,1}$ or $\delta_{1,--}$. 
It is also noted that the energy gap between $|\Psi_{++}\rangle$ and $|\Psi_{--}\rangle$ is invariant under the perturbation, i.e., $\Lambda_{++}-\Lambda_{--}=\delta_{++,2}+\delta_{2,--}=\delta_{++,1}+\delta_{1,--}=\eta$.}
\label{fig:schematic_crossResonantDrive}
\end{figure}

\begin{figure}
\centering
  \includegraphics[width=0.4\columnwidth]{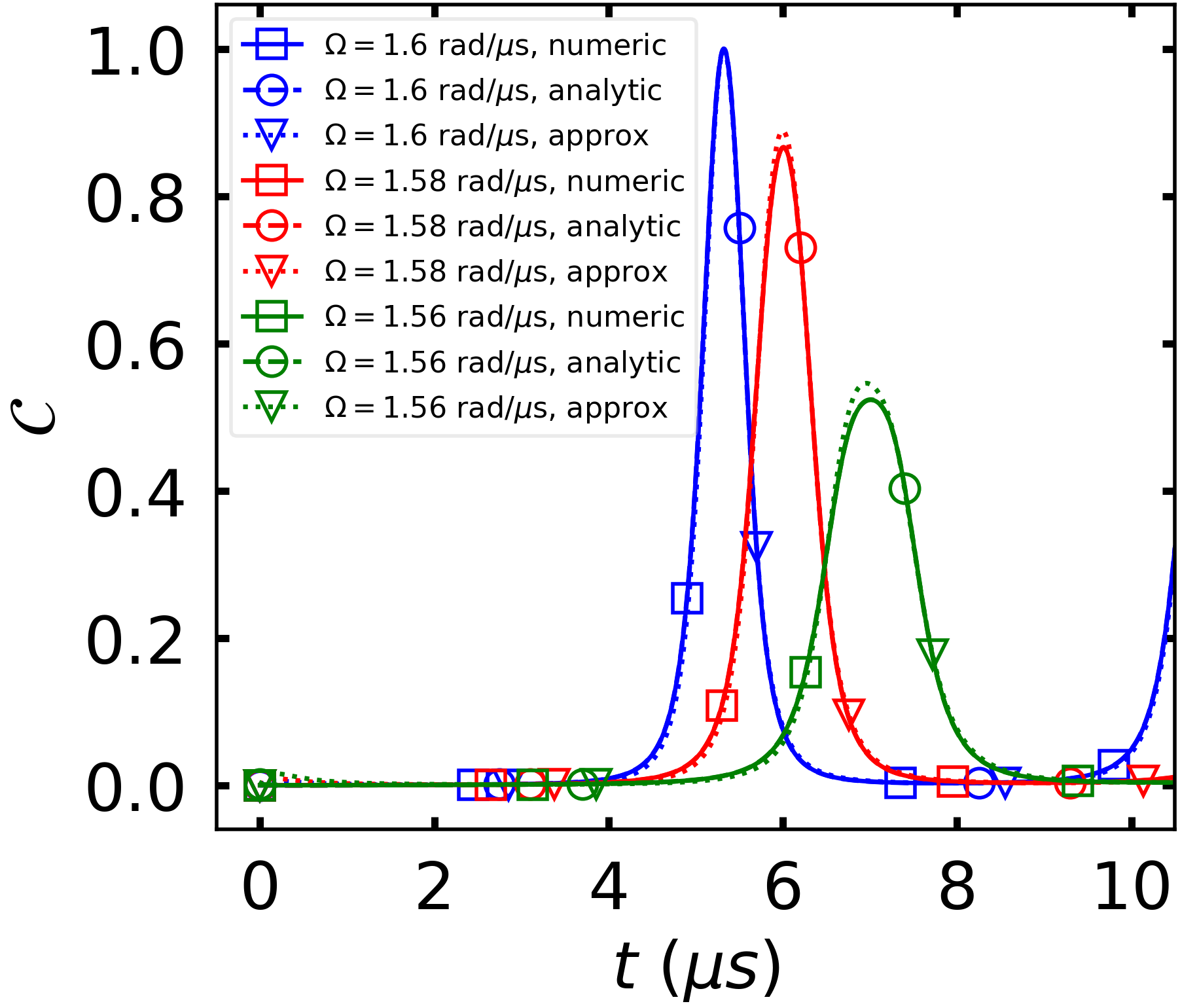} 
\caption{
Comparison of the calculated concurrence ${\cal C}$ as a function of time from the first-order non-Hermitian perturbation theory [circles on dashed lines, see Eq. (\ref{eq:C_analytics})] and the approximate analytical expression [down-pointing triangles on dotted lines, see Eq. (\ref{eq:C_approx})] with the fully numerical calculation (squares on solid lines) at three $\Omega$ values in the ${\cal PT}$-symmetric phase: $\Omega=1.6$ rad/$\mu$s (blue), $\Omega=1.58$ rad/$\mu$s (red), and $\Omega=1.56$ rad/$\mu$s (green). The initial state is $|\psi(0)\rangle=|ff\rangle$, and other parameters are: $\gamma=6\,\mu\rm{s}^{-1}$ and $J=0.001$ rad/$\mu$s.} 
\label{fig:analytics_approx_numerics}
\end{figure}

In order to obtain readable analytical expressions, 
we further assume the overlap of non-degenerate and degenerate subspaces is vanishingly small, which essentially comes from the weak inter-qubit coupling $J$ (i.e., $J\ll\gamma$) under our consideration. 
This leads to a further simplification of the constructed basis states, i.e., Eqs.~(\ref{eq:Psi_--})-
(\ref{eq:barPsi_++}) and (\ref{eq:Psi_1})-
(\ref{eq:barPsi_2}), of the full Hamiltonian, namely
\begin{eqnarray}
|\Psi_{--}\rangle &\approx& |\tilde{\psi}_{--}\rangle 
+ \frac{\langle \tilde{\bar{\psi}}_{++}| H_{\rm int} |\tilde{\psi}_{--}\rangle}{\lambda_{--}-\lambda_{++}} |\tilde{\psi}_{++}\rangle 
=
|\tilde{\psi}_{--}\rangle 
+\frac{8J\Omega^2}{\eta^3} |\tilde{\psi}_{++}\rangle 
=
\frac{1}{2\eta^4}
\begin{pmatrix}
\eta^3(-i\gamma + \eta) + 8J\Omega^2(i\gamma+\eta) \\
- 4\Omega (\eta^3-8J\Omega^2) \\
- 4\Omega (\eta^3-8J\Omega^2) \\
\eta^3(i\gamma + \eta) + 8J\Omega^2(-i\gamma+\eta)
\end{pmatrix} , 
\label{eq:Psi_--_Approx}
\end{eqnarray}

\begin{eqnarray}
|\bar{\Psi}_{--}\rangle &\approx& |\tilde{\bar{\psi}}_{--}\rangle 
+ \frac{\langle \tilde{\psi}_{++}| H_{\rm int} |\tilde{\bar{\psi}}_{--}\rangle}{\bar{\lambda}_{--}-\bar{\lambda}_{++}} |\tilde{\bar{\psi}}_{++}\rangle 
=
|\tilde{\bar{\psi}}_{--}\rangle 
+ \frac{8J\Omega^2}{\eta^3} |\tilde{\bar{\psi}}_{++}\rangle 
=
\frac{1}{2\eta^4}
\begin{pmatrix}
\eta^3(i\gamma + \eta) + 8J\Omega^2(-i\gamma+\eta) \\
- 4\Omega (\eta^3-8J\Omega^2) \\
- 4\Omega (\eta^3-8J\Omega^2) \\
\eta^3(-i\gamma + \eta) + 8J\Omega^2(i\gamma+\eta)
\end{pmatrix} , 
\label{eq:barPsi_--_Approx}
\end{eqnarray}

\begin{eqnarray}
|\Psi_{++}\rangle &\approx& |\tilde{\psi}_{++}\rangle 
+ \frac{ \langle \tilde{\bar{\psi}}_{--}| H_{\rm int} |\tilde{\psi}_{++}\rangle}{\lambda_{++}-\lambda_{--}} |\tilde{\psi}_{--}\rangle 
=
|\tilde{\psi}_{++}\rangle 
- \frac{8J\Omega^2}{\eta^3} |\tilde{\psi}_{--}\rangle
=
\frac{1}{2\eta^4}
\begin{pmatrix}
\eta^3(i\gamma + \eta) + 8J\Omega^2(i\gamma-\eta) \\
 4\Omega (\eta^3+8J\Omega^2) \\
 4\Omega (\eta^3+8J\Omega^2) \\
\eta^3(-i\gamma + \eta) - 8J\Omega^2(i\gamma+\eta)
\end{pmatrix} , 
\label{eq:Psi_++_Approx}
\end{eqnarray}

\begin{eqnarray}
|\bar{\Psi}_{++}\rangle &\approx& |\tilde{\bar{\psi}}_{++}\rangle 
+ \frac{ \langle \tilde{\psi}_{--}| H_{\rm int} |\tilde{\bar{\psi}}_{++}\rangle}{\bar{\lambda}_{++}-\bar{\lambda}_{--}} |\tilde{\bar{\psi}}_{--}\rangle 
=
|\tilde{\bar{\psi}}_{++}\rangle
- \frac{8J\Omega^2}{\eta^3} |\tilde{\bar{\psi}}_{--}\rangle 
=
\frac{1}{2\eta^4}
\begin{pmatrix}
\eta^3(-i\gamma + \eta) - 8J\Omega^2(i\gamma+\eta) \\
 4\Omega (\eta^3+8J\Omega^2) \\
 4\Omega (\eta^3+8J\Omega^2) \\
\eta^3(i\gamma + \eta) + 8J\Omega^2(i\gamma-\eta)
\end{pmatrix} , 
\label{eq:barPsi_++_Approx}
\end{eqnarray}
and
\begin{eqnarray}
|\Psi_{1} \rangle &=& |\tilde{\psi}_1\rangle 
= \frac{1}{\sqrt{2}} (|\tilde{\psi}_{-+}\rangle - |\tilde{\psi}_{+-}) , \label{eq:Psi_1_Approx} \\
|\bar{\Psi}_1\rangle &=& |\tilde{\bar{\psi}}_1\rangle 
= \frac{1}{\sqrt{2}} (|\tilde{\bar{\psi}}_{-+}\rangle -|\tilde{\bar{\psi}}_{+-}\rangle) , \label{eq:barPsi_1_Approx}\\
|\Psi_{2} \rangle &\approx& |\tilde{\psi}_2\rangle 
= \frac{1}{\sqrt{2}} (|\tilde{\psi}_{-+}\rangle + |\tilde{\psi}_{+-}), \label{eq:Psi_2_Approx} \\
|\bar{\Psi}_2\rangle &\approx& |\tilde{\bar{\psi}}_2\rangle 
= \frac{1}{\sqrt{2}} (|\tilde{\bar{\psi}}_{-+}\rangle + |\tilde{\bar{\psi}}_{+-}\rangle). 
\label{eq:barPsi_2_Approx}
\end{eqnarray}
An illustration of the approximate perturbed energy levels, i.e., the energies of the states $|\Psi_{++}\rangle$, $|\Psi_{--}\rangle$, $|\Psi_{1}\rangle$, and $|\Psi_{2}\rangle$ in the Hilbert space, is presented in Fig. \ref{fig:schematic_crossResonantDrive} where the unperturbed levels are included as grey dashed lines.

\subsection{Analytical results for the entangled state derived from initial state $|ff\rangle$}

We substitute the approximate basis states in Eqs. (\ref{eq:Psi_--_Approx})-(\ref{eq:barPsi_2_Approx}) into Eqs. (\ref{eq:stateNormlized}) with the initial condition $|\psi(0)\rangle=|ff\rangle$ and then obtain an approximate state $|\tilde{\psi}(t)\rangle=\frac{|\psi(t)\rangle}{| |\psi(t)\rangle |}$ with
\begin{eqnarray}
|\psi(t)\rangle 
&=&
\frac{\eta^3(i\gamma + \eta) + 8J\Omega^2(i\gamma-\eta)}{2\eta^4} 
e^{-it\Lambda_{++}} | \Psi_{++}\rangle
+ \frac{\eta^3(-i\gamma + \eta) + 8J\Omega^2(i\gamma+\eta)}{2\eta^4}  
e^{-it\Lambda_{--}} | \Psi_{--}\rangle \notag\\
&&
- \frac{4\Omega(\eta^2+2iJ\gamma)}{\sqrt{2}\eta^3} 
e^{-it\Lambda_{2}} | \Psi_{2}\rangle ,
\end{eqnarray}
which becomes $ |\psi(t)\rangle = (\alpha^{\prime},\beta^{\prime},\zeta^{\prime},\delta^{\prime})^T$ in the basis $\{|ff\rangle, |fe\rangle, |ef\rangle, |ee\rangle\}$, where
\begin{eqnarray}
\alpha^{\prime} &\approx& \frac{1}{16\eta^8}
e^{-\frac{t}{2} [\gamma +i\eta + iJ (1+\frac{\gamma^2}{\eta^2})]} 
\Big\{ 4(e^{it\eta}-1)J\eta^3(\gamma^2+\eta^2)^2 +J^2(\gamma^2+\eta^2)^2[(\eta-i\gamma)^2 +e^{it\eta}(\eta+i\gamma)^2] \notag\\
&& +4\eta^6 \Big[(\eta^2-\gamma^2)(e^{it\eta} +1) -2i\gamma\eta (e^{it\eta} -1)  + 2(\eta^2+\gamma^2)e^{\frac{it}{2} [\eta +J (1+3\frac{\gamma^2}{\eta^2})]} \Big]
\Big\} , \\
\beta^{\prime}=\zeta^{\prime} &\approx& \frac{\Omega}{4\eta^8}
e^{-\frac{t}{2} [\gamma +i\eta + iJ (1+\frac{\gamma^2}{\eta^2})]}
\Big\{ 4\eta^6 [i\gamma (1+e^{it\eta} -2e^{\frac{it}{2} [\eta +J(1 +3\frac{\gamma^2}{\eta^2})]}) -\eta(e^{it\eta}-1)]  \notag\\
&& + J^2(\gamma^2+\eta^2)^2[(\eta+i\gamma)e^{it\eta} -\eta+i\gamma] + 4iJ\gamma\eta^3 (\gamma^2+\eta^2)(1-e^{it\eta})
\Big\} , \\
\delta^{\prime} &\approx& \frac{4\Omega^2}{\eta^8}
e^{-\frac{t}{2} [\gamma +i\eta + iJ (1+\frac{\gamma^2}{\eta^2})]} 
\Big\{ \eta^6 [1+e^{it\eta} -2e^{\frac{it}{2} [\eta + J(1+3\frac{\gamma^2}{\eta^2})]}] +J\eta^3(\eta^2-\gamma^2)(e^{it\eta}-1) +64J^2\Omega^4(e^{it\eta}+1)
\Big\} . \notag\\
\end{eqnarray}
In terms of the Bell states in Eqs.~(\ref{eq:Bell_++--}) and (\ref{eq:Bell_-++-}), this normalized state $ |\psi(t)\rangle$ can be also written as $|\psi(t)\rangle= \frac{\alpha^{\prime}+\delta^{\prime}}{\sqrt{2}} |e_1\rangle + i\frac{\alpha^{\prime}-\delta^{\prime}}{\sqrt{2}} |e_2\rangle + i\frac{\beta^{\prime}+\zeta^{\prime}}{\sqrt{2}} |e_3\rangle + \frac{\beta^{\prime}-\zeta^{\prime}}{\sqrt{2}} |e_4\rangle$, 
and therefore the concurrence in Eq. (\ref{eq:C_analytics}) becomes
\begin{eqnarray}
C &\approx& 2\Big| \frac{ {\cal A} }{ {\cal B} } \Big| ,
\label{eq:C_approx}
\end{eqnarray}
where
\begin{eqnarray}
{\cal A} &=& \eta^2(\eta^2+\gamma^2)\left[ e^{itJ\left(1+3\frac{\gamma^2}{\eta^2}\right) } -1\right] ,
\end{eqnarray}
and
\begin{eqnarray}
{\cal B} &=& 
\gamma^2(\gamma^2-\eta^2)(1+e^{2it\eta}) 
+ 2e^{it\eta}\Big\{(\gamma^2+\eta^2)(3\gamma^2+2\eta^2) \notag\\
&&-4\gamma(\gamma^2+\eta^2)\cos\left[\frac{tJ}{2}\left(1+3\frac{\gamma^2}{\eta^2} \right) \right]\left[\gamma\cos\left(\frac{t\eta}{2}\right) -\eta\sin\left(\frac{t\eta}{2}\right)\right] -2\gamma^3\eta\sin(t\eta) \Big\}
\end{eqnarray}
with $\eta=\sqrt{16\Omega^2-\gamma^2}>0$. This is Eq.~(3) of the main text. 
When $J=0$, it is easy to confirm that ${\cal A}(J=0)=0$, 
indicating that the two qubits are not entangled. In Fig. \ref{fig:analytics_approx_numerics}, we compare the approximate result for the concurrence $\cal C$ as a function of time from Eq. (\ref{eq:C_approx}) with both the analytic result from Eq. (\ref{eq:C_analytics}) and the full numerical calculations, all of which show good agreement.


\begin{figure}[htp]
\centering
  \includegraphics[width=0.46\columnwidth]{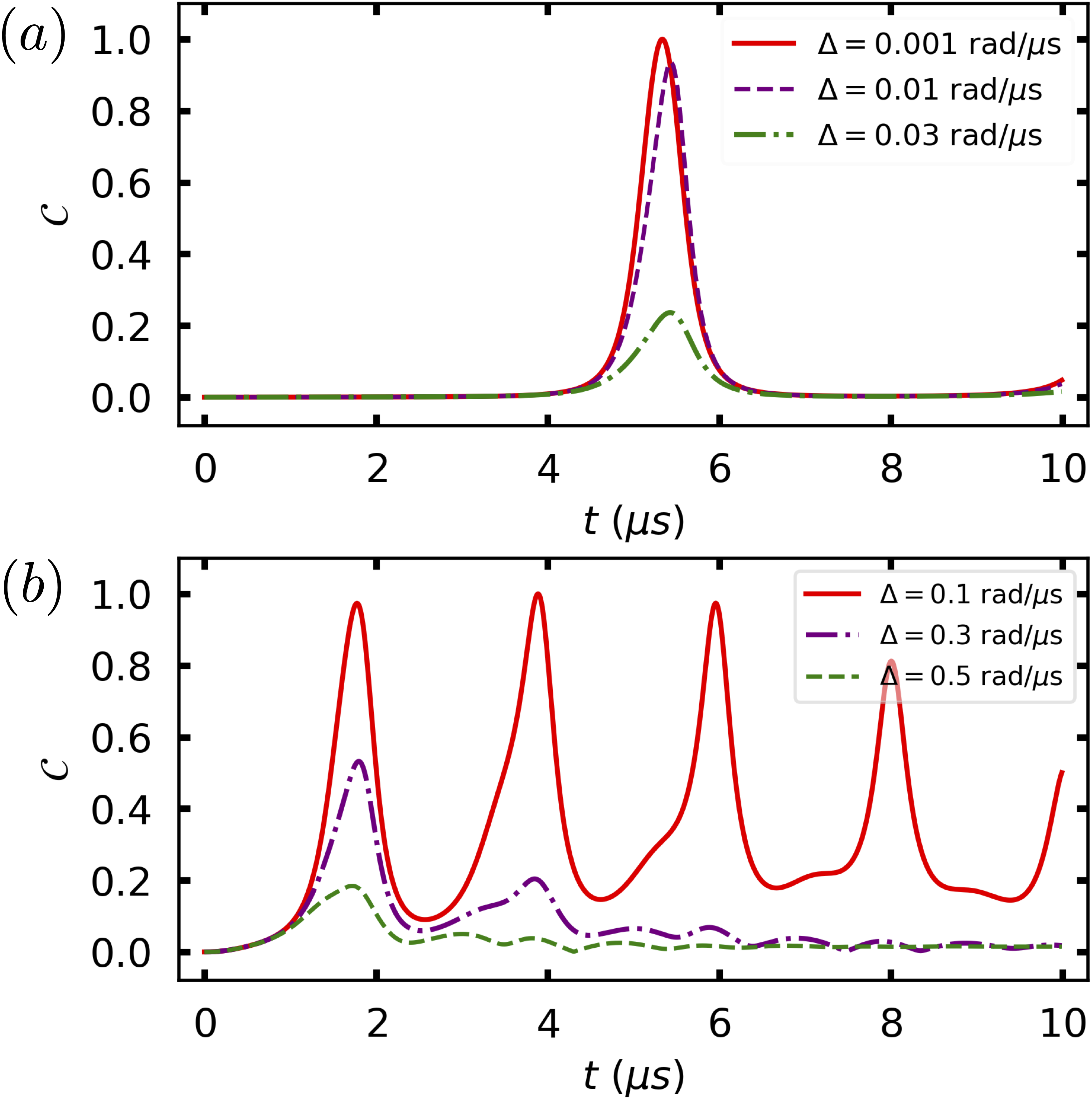} 
\caption{
{\bf Time evolution of concurrence under nonresonant qubit drives}. The qubit drives are assumed to have the same detuning, i.e., $\Delta_1 = \Delta_2 \equiv \Delta$. Two $J$ values are considered: $J=0.001$ rad/$\mu$s (a) and $J=0.1$ rad/$\mu$s (b), with corresponding qubit drives $\Omega$ equal to $1.6$ rad/$\mu$s and $2.2$ rad/$\mu$s, respectively. The initial state $|\psi(t=0)\rangle=|ff\rangle$, and other parameters used are $\gamma_1=\gamma_2=6$ $\mu{\rm s}^{-1}$. The fourth-order EP is located at $\Omega=1.5$ rad/$\mu$s for these parameters.} 
\label{fig:FigS_offresonance} 
\end{figure}

\section{Off-resonant drive}

In the main text, we consider two non-Hermitian qubits under resonant drives. Here we present the effect of non-vanishing detuning ($\Delta_1, \Delta_2\neq 0$) at two $J$ values, 
namely, $J=0.001$ rad/$\mu$s in Fig.~\ref{fig:FigS_offresonance}(a) and
$J=0.1$ rad/$\mu$s in Fig.~\ref{fig:FigS_offresonance}(b). 
For simplicity, we assume $\Delta_1=\Delta_2 \equiv\Delta$. We see that for both $J$ values, while the concurrence decreases for non-zero detuning, the maximal value of the concurrence is still close to one when the detuning is of equal magnitude to the coupling, i.e., when $\Delta=J$. Furthermore, we find that the larger $J$ value provides stronger tolerance to finite detuning, i.e., with larger $J$ one can still achieve high concurrence for larger values of $\Delta$ [compare the maxima in Fig.~\ref{fig:FigS_offresonance}(b) with the maxima in Fig.~\ref{fig:FigS_offresonance}(a)]. 
This can be understood physically as a result of a larger $J$ coupling requiring a larger qubit drive $\Omega$ to reach the maximal entanglement [Fig.~3(c) of the main text] when the system is farther from the fourth-order EP. Because of the larger value of $\Omega$, this system can show more robustness to the detuning.

\begin{figure}
\centering
  \includegraphics[width=0.46\columnwidth]{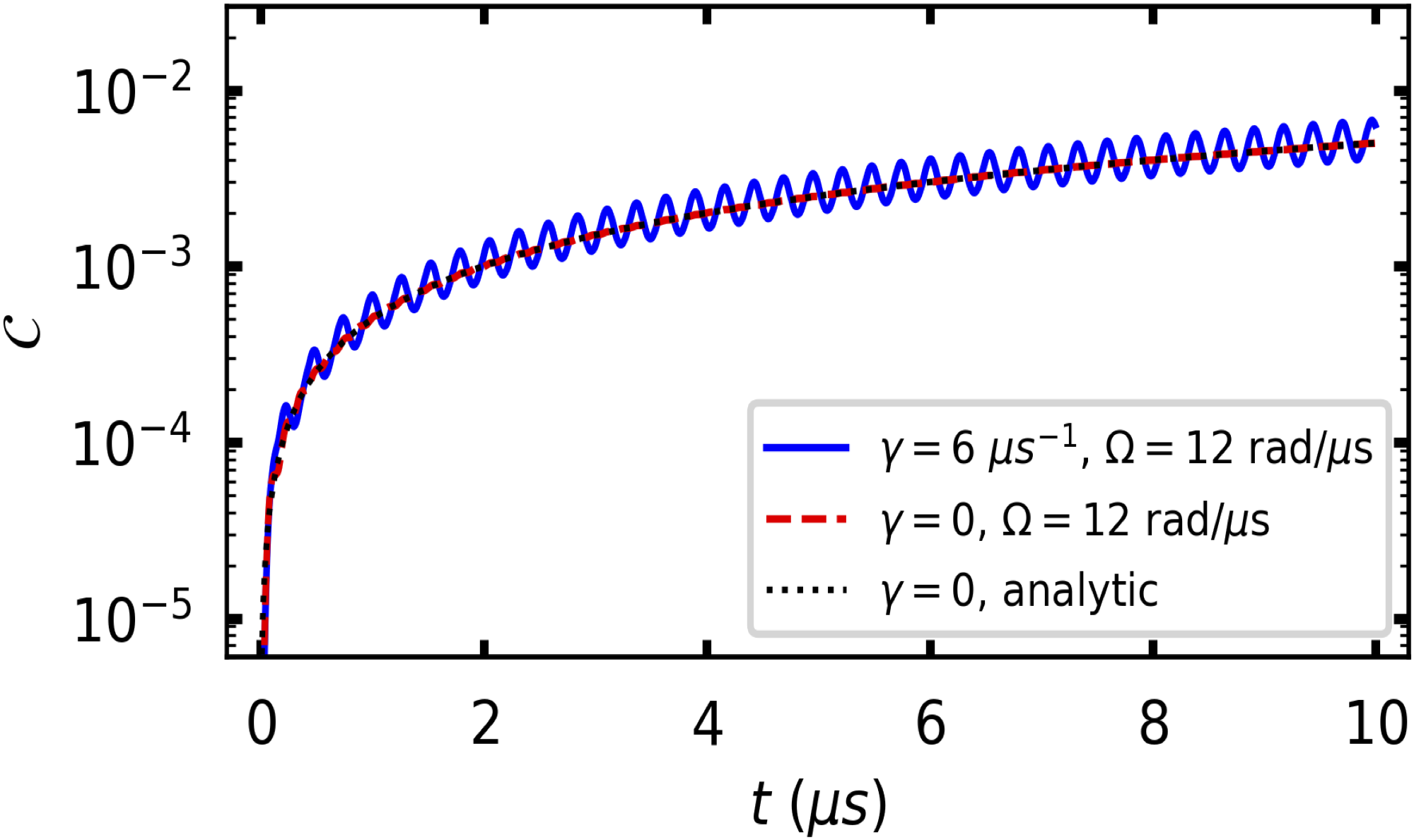} 
\caption{{\bf Strong driving and the Hermitian limit}. Time evolution of concurrence for non-Hermitian qubits in the strong driving regime $\Omega \gg \Omega_{\rm{EP}}$ (blue solid curve). The Hermitian limit ($\gamma =0$) is provided for comparison (red dashed curve). All plots are calculated with $|\psi(t=0)\rangle=|ff\rangle$ and $J=10^{-3}$ rad/$\mu$s. 
With this initial state and in this regime of strong driving in the ${\cal PT}$-symmetric phase, the concurrence for the perturbatively coupled Hermitian qubits can be calculated from ${\cal C}=\sin(\frac{tJ}{2})$ (Eq.~(3) of the main text) and is independent of drive strength $\Omega$ (black dotted curve).} 
\label{fig:FigS_nonHermitianReturn}
\end{figure}

\section{Hermitian limit}

In the Hermitian limit (i.e., $\gamma=0$), when starting from initial state $|ff\rangle$ and evolving under strong driving $\Omega \gg \Omega_{\rm EP}$, the concurrence is given by ${\cal C} =|\sin(\frac{tJ}{2})|$, obtained from Eq.~(3) of the main text. The entanglement generation is then independent of the drive amplitude $\Omega$. This is illustrated by comparison between the red dashed and black dotted lines of Fig.~\ref{fig:FigS_nonHermitianReturn}. The blue solid line in Fig.~\ref{fig:FigS_nonHermitianReturn} show the concurrence for non-Hermitian qubits under such strong driving with $\Omega \gg \Omega_{\rm EP}$. This oscillates around the red dashed line, showing that the concurrence is well-approximated by the Hermitian limit in this regime. In particular, when $J$ is weak, as considered in this work, the qubits are approximately decoupled and the structure of the eigenstates is accordingly symmetric (see, e.g., the dashed grey lines in Fig.~\ref{fig:schematic_crossResonantDrive}). Consequently the entanglement, which is now independent of the dynamics of the individual qubits, 
grows very slowly on the time scale $\sim1/J (\gg 1/\Omega)$. This behavior is similar to that of a Hermitian system under driving and is rationalized below. 

\begin{figure}[htp]
\centering
  \includegraphics[width=0.89\columnwidth]{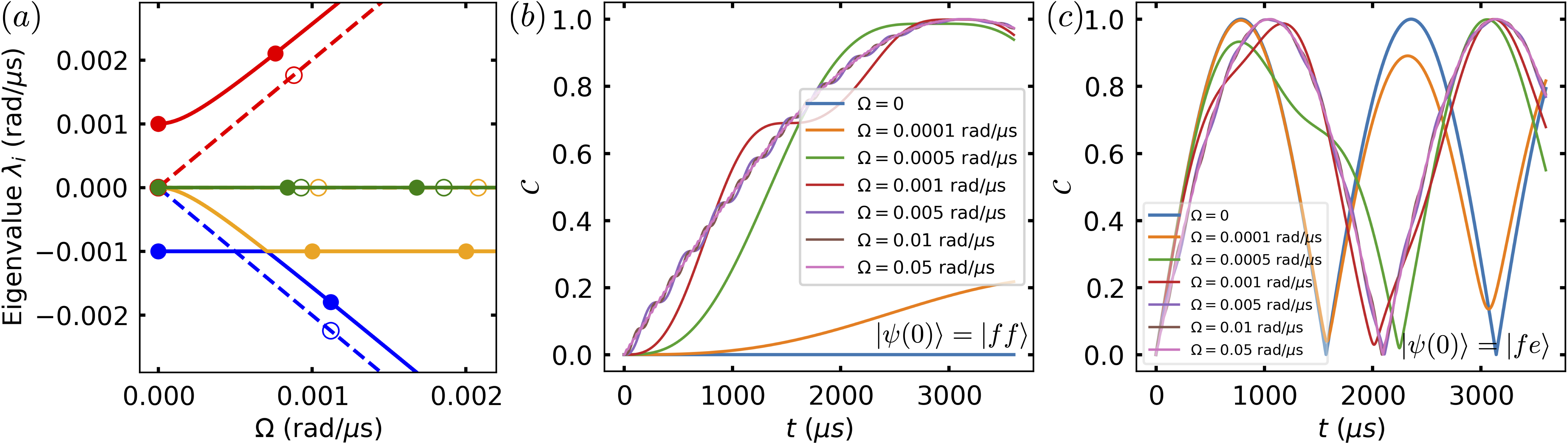} 
\caption{
{\bf Entanglement generation near a Hermitian degeneracy point}. (a) Dependence of the eigenvalues of two coupled Hermitian qubits on the qubit drive strength $\Omega$ at two $J$ values: $J=0$ (dashed curves and empty circles) and $J=0.001$ rad/$\mu$s (solid curves and full circles). The two qubits have the same drive amplitude $\Omega_1=\Omega_2 \equiv \Omega$. 
The four dashed lines meet at $J=0$, signifying degeneracy of all eigenvalues here. The circles are virtual aids to distinguish the solid and dashed lines. 
(b) Concurrence evolution near a Hermitian degeneracy point at different drive amplitudes $\Omega$ for initial state  $|\psi(t=0)\rangle=|ff\rangle$. 
(c) Concurrence evolution near a Hermitian degeneracy point at different drive amplitudes $\Omega$ for initial state $|\psi(t=0)\rangle=|fe\rangle$. Panels (b) and (c) both use parameter values $J=0.001$ rad/$\mu$s, $\Delta_1=\Delta_2=0$, and $\gamma_1=\gamma_2=0$.} 
\label{fig:FigS_Hermitian_DP}
\end{figure}

The system can possess a Hermitian degeneracy at the Hermitian limit, where the eigenvalues are equal but the eigenvectors are linearly independent. We now discuss entanglement generation near such a Hermitian degeneracy point and show that this differs significantly from the generation near an EP. Fig.~\ref{fig:FigS_Hermitian_DP}(a) summarizes the dependence of eigenvalues on the qubit drive $\Omega$ and the inter-qubit coupling $J$. It is evident that there exists a Hermitian degeneracy when $J=0$ and $\Omega=0$ (see dashed curves and empty circles). Nonzero values of $\Omega$ or $J$ can lift this degeneracy, as indicated by the solid curves with full circles.

We now consider the time evolution of concurrence near the Hermitian degeneracy point at two different initial states $|ff\rangle$ and $|fe\rangle$, where the former is used in our current work and the latter is conventionally used to establish entanglement through energy swap without a single qubit drive~\cite{MajerSchoelkopf07nature}. For the initial state $|ff\rangle$, Fig.~\ref{fig:FigS_Hermitian_DP}(b) shows that the time scale (a few milliseconds) for entanglement generation in the Hermitian system is significantly longer than that near the EP in the non-Hermitian system (a few microseconds). 
We interpret this result as follows: the qubit drive $\Omega$ drives the initial state $|ff\rangle$ into subspace $\{|fe\rangle, |ef\rangle \}$, and then the entanglement is established at a time scale of $1/J$, similar to that of the conventional energy swap mechanism. However, the concurrence remains zero when $\Omega=0$ (blue curve), since without any driving the two-qubit state remains in the $|ff\rangle$ state.

For the initial state $|fe\rangle$, Fig.~\ref{fig:FigS_Hermitian_DP}(c) shows that the time scale (one millisecond) for entanglement generation  is shorter than that
for the initial state $|ff\rangle$, but that it is still significantly longer than the time scale of a few microseconds when operating near the EP in the non-Hermitian system. Note that for this initial state, there is still entanglement generation when $\Omega=0$, consistent with the energy swap mechanism. 

These results for entanglement generation near a degeneracy of a Hermitian system imply that our observation of significantly enhanced rates of entanglement generation near an EP does not have any analogy in the case of Hermitian degeneracy.
We emphasize that the enhancement of entanglement generation for the non-Hermitian qubits is due to proximity to an EP and there is no such enhancement for Hermitian qubits when close to a conventional degeneracy point. 
We note further that adding a drive to the Hermitian case with the initial state $|fe\rangle$ also does not enhance the rate of entanglement generation. 

\begin{figure}[htp!]
\centering
  \includegraphics[width=0.46\columnwidth]{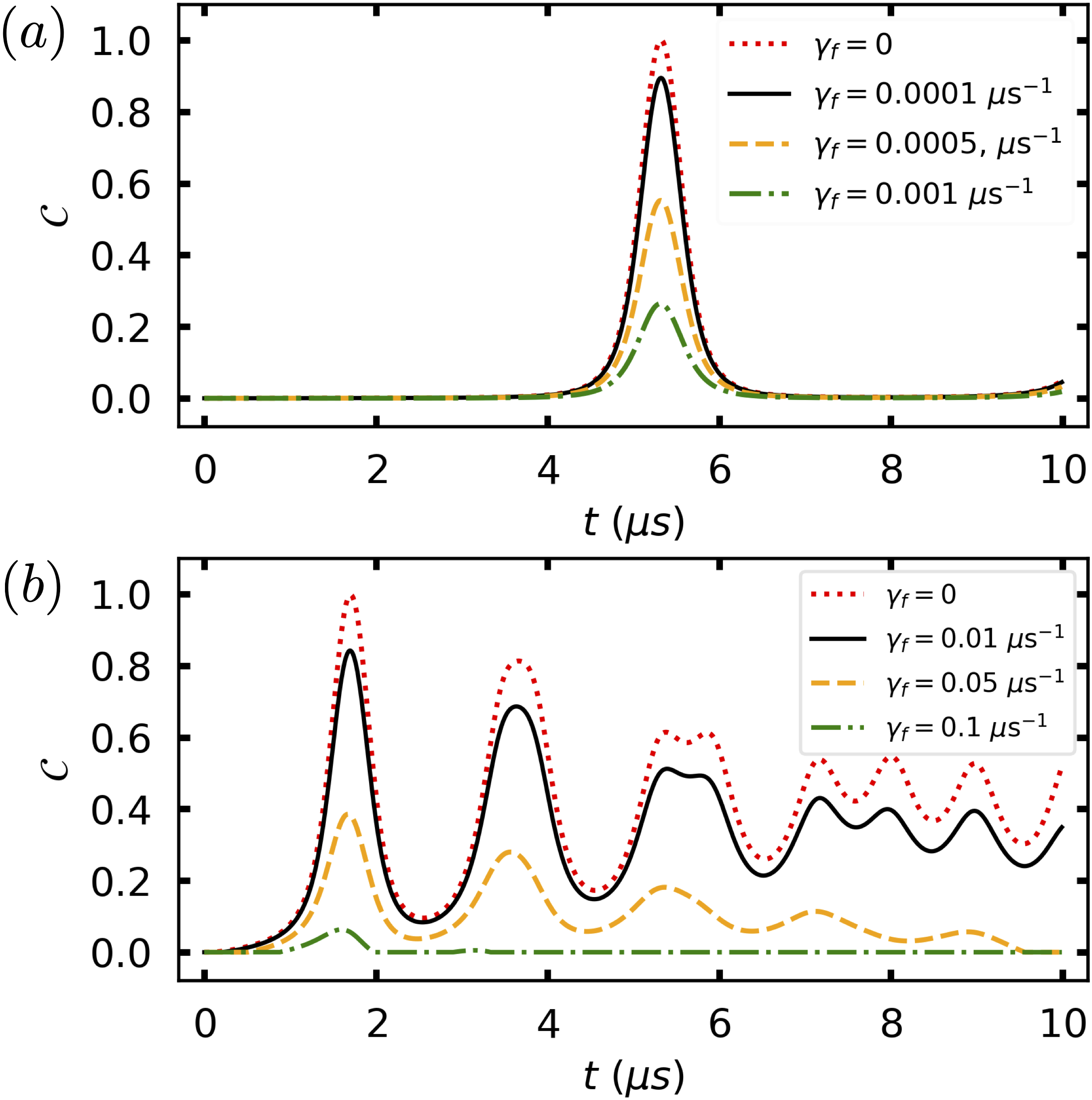} 
\caption{
Time evolution of concurrence under decoherence from spontaneous emission of the $|f\rangle$ states of the two non-Hermitian qubits with the initial state $|\psi(t=0)\rangle=|ff\rangle$, for two values of $J$ with corresponding qubit drive strengths $\Omega$. 
The two qubits have the same $|f\rangle$ state decay rate $\gamma_f$.
(a) $J=0.001$ rad/$\mu$s and $\Omega=1.6$ rad/$\mu$s. (b) $J=0.1$ rad/$\mu$s and $\Omega=2.2$ rad/$\mu$s.
 Calculations are presented in each panel for three $\gamma_f$ values: $\gamma_f=\{0.1J,\, 0.5J,\, J\}$. Results for $\gamma_f=0$ are also included for reference. Other parameters used are $\Delta=0$ and $\gamma=6$ $\mu{\rm s}^{-1}$.} 
\label{fig:FigS_entanglement_Omega_Lindblad}
\end{figure}

\begin{figure}
\centering
  \includegraphics[width=0.46\columnwidth]{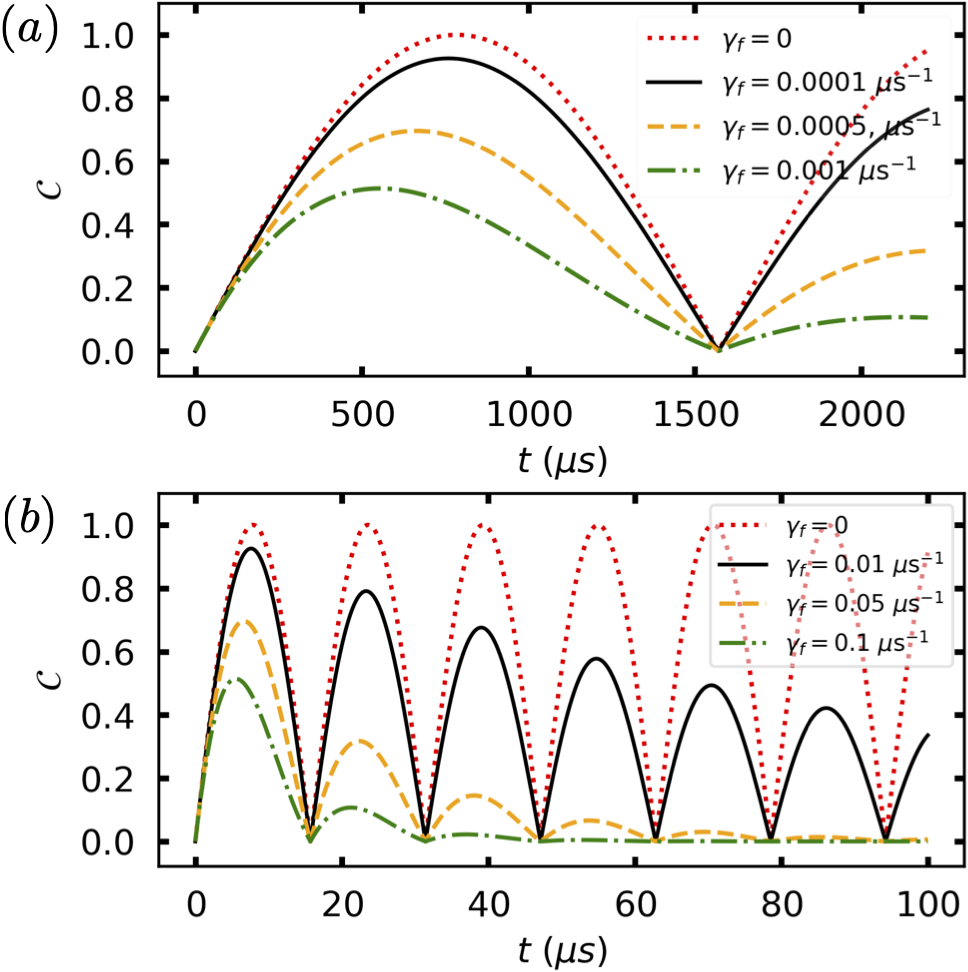} 
\caption{
Time evolution of concurrence under decoherence from spontaneous emission of the $|f\rangle$ states of the two Hermitian qubits with the initial state $|\psi(t=0)\rangle=|fe\rangle$, for two values of $J$ without qubit drive. (a) $J=0.001$ rad/$\mu$s. (b) $J=0.1$ rad/$\mu$s.
The two qubits have the same $|f\rangle$ state decay rate $\gamma_f$. Calculations are presented in each panel for three $\gamma_f$ values: $\gamma_f=\{0.1J,\, 0.5J,\, J\}$. Results for $\gamma_f=0$ are also included for reference. Other parameters used are $\Delta=\Omega=\gamma=0$.}
\label{fig:FigS_entanglement_Omega_Lindblad_gamma_e=0_Omega=0}
\end{figure}


\section{Effect of decoherence }

In addition to the dissipative decay $\gamma$ of the lower level $|e\rangle$ of each qubit that is taken into account in the definition of the non-Hermitian Hamiltonian, we can also consider the spontaneous emission from the upper level state $|f\rangle_i$ ($i=1, 2$) at $\gamma_{i,f}$ as a source of decoherence affecting the entanglement generation. The relevant physics for a single non-Hermitian qubit case undergoing such additional decoherence has been studied in~\cite{ChenMurch21prl}. 
To reveal the effect of this decoherence we consider the Lindblad master equation
\begin{eqnarray}
\frac{d\rho}{dt} &=& -i\left(H\rho-\rho H^{\dagger} \right) 
+\sum_{i=1, 2} \gamma_{i,f} \left(\sigma_i^{-}\rho\sigma_i^{+} 
 -\frac{1}{2}\left\{\sigma_i^{+}\sigma_i^{-}, \rho \right\}\right) 
\label{eq:ME}
\end{eqnarray}
with $H$ given in Eq.~(2) of the main text and Lindblad operators $\sigma_i^{-}=|g\rangle_i\langle e|$ and $\sigma_i^+ = |e\rangle_i\langle g| \equiv (\sigma_i^{-})^\dagger$. 
With the two-qubit density operator $\rho$ from Eq.~(\ref{eq:ME}), the entanglement between the two non-Hermitian qubits then can be measured via the mixed state concurrence~\cite{Wootters98prl} 
\begin{eqnarray}
{\cal C} = \max(0,\tau_1-\tau_2-\tau_3-\tau_4) ,
\end{eqnarray}
where $\tau_1, \tau_2, \tau_3, \tau_4$ are the eigenvalues, in decreasing order, of the Hermitian matrix $R = \sqrt{\sqrt{\rho} \tilde{\rho}\sqrt{\rho}}$ with  $\tilde{\rho} = (\sigma_y\otimes\sigma_y)\rho^* (\sigma_y\otimes\sigma_y)$. 

For simplification, we assume the two qubits have the same decay rate for their $|f\rangle_i$ states, i.e., $\gamma_{1,f}=\gamma_{2,f} \equiv \gamma_f$. The concurrence evolution for two values of $J$ ($0.001$ rad/$\mu$s and $0.1$ rad/$\mu$s) are presented in Fig.~\ref{fig:FigS_entanglement_Omega_Lindblad}, for three values of decoherence $\gamma_f$ that are scaled to $J$ as $\gamma_f=\{0.1J,\, 0.5J,\, J\}$ under the initial state $|\psi(t=0)\rangle=|ff\rangle$. 
Panel (a) shows the dynamics for $J=0.001$ rad/$\mu$s with the optimal drive $\Omega=1.6$ rad/$\mu$s for this $J$ value (see Fig.~\ref{fig:FigS_optimal_Omega_T}). Panel (b) shows the dynamics for the significantly larger qubit coupling $J=0.1$ rad/$\mu$s, for which the optimal drive amplitude is $\sim \Omega \simeq 2.2$ rad/$\mu$s (see Fig.~\ref{fig:FigS_large_J}(a)). 
We see that adding the decoherence only slightly changes the time for reaching the maximal concurrence. However it does reduce the value of maximal concurrence. Not surprisingly, the reduction is particularly severe for the largest decoherence rates that are equal to the qubit coupling, $\gamma_f=J$. 
Fig.~\ref{fig:FigS_entanglement_Omega_Lindblad_gamma_e=0_Omega=0} shows the concurrence evolution of the corresponding Hermitian systems calculated with the same dissipation rates and the same initial state.  
Comparing Fig.~\ref{fig:FigS_entanglement_Omega_Lindblad} with Fig.~\ref{fig:FigS_entanglement_Omega_Lindblad_gamma_e=0_Omega=0} it is evident that the maximal value of ${\cal C}$ is somewhat lower than that for the Hermitian systems when calculated with the same dissipation rates from the $|f\rangle$ states, but the time scale for entanglement generation is still considerably shorter for the non-Hermitian system.

\begin{figure}[htp!]
\centering
  \includegraphics[width=0.46\columnwidth]{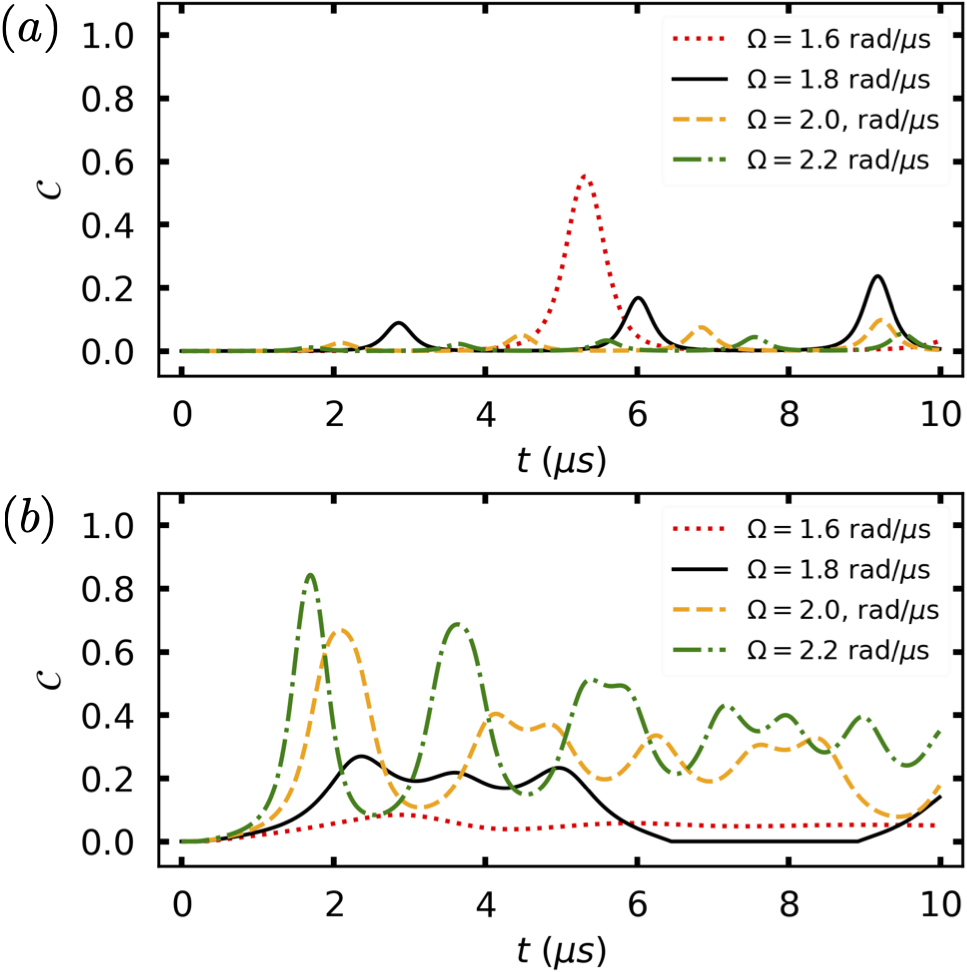} 
\caption{
Time evolution of concurrence under decoherence from spontaneous emission of the $|f\rangle$ states of the two non-Hermitian qubits with the initial state $|\psi(t=0)\rangle=|ff\rangle$, for two values of $J$ and for different values of qubit drive strength $\Omega$. (a) $J=0.001$ rad/$\mu$s and $\gamma_f=0.0005$ $\mu$s$^{-1}$. (b) $J=0.1$ rad/$\mu$s and $\gamma_f=0.01$ $\mu$s$^{-1}$.
The two qubits have the same $|f\rangle$ state decay rate $\gamma_f$. Calculations are presented in each panel for four $\Omega$ values: $\Omega=\{1.6,\, 1.8,\, 2.0,\, 2.2\}$ rad/$\mu$s. Other parameters used are $\Delta=0$ and $\gamma=6$ $\mu{\rm s}^{-1}$.} 
\label{fig:FigS_entanglement_gamma_f_Lindblad_varying_Omega}
\end{figure}

We note that for a fixed value of $\gamma_f$, e.g., $\gamma_f=0.0005$ $\mu$s$^{-1}$ in Fig.~\ref{fig:FigS_entanglement_gamma_f_Lindblad_varying_Omega}(a), increasing the drive amplitude beyond $\Omega^*=\Omega=1.6$ rad/$\mu$s for $J=0.001$ rad/$\mu$s also reduces the rate of entanglement generation for the non-Hermitian systems. This is understandable since the drive increases the population at $|f\rangle$ state with $\gamma_f=0.0005$ $\mu$s$^{-1}$ and therefore effectively enhances decoherence. 
For the larger coupling strength, $J=0.1$ rad/$\mu$s with the corresponding optimal drive  $\sim \Omega^*=\Omega\simeq 2.2$ rad/$\mu$s [see Fig.~\ref{fig:FigS_large_J}(a)], Fig.~\ref{fig:FigS_entanglement_gamma_f_Lindblad_varying_Omega}(b) shows for a fixed value of $\gamma_f=0.01$ $\mu$s$^{-1}$, decreasing the drive amplitude away from $\Omega^*$ {\it toward} the EP (i.e., $\Omega_{\rm EP}<\Omega<\Omega^*$) reduces the concurrence. This suggests an enhancement of the effect of decoherence on the system when it is close to the EP as discussed in Ref.~\cite{ChenMurch21prl}, 
although more detailed analysis is required to determine the scaling of this effect. 
Overall, our results demonstrate that maximal concurrence is achieved at a location $\Omega^*$  that is close to but not equal to the value at the EP, i.e., $\Omega_{\rm EP}$, suggesting there is a competition between EP-enhanced entanglement and EP-enhanced decoherence in the range $\Omega_{\rm EP}<\Omega^*<\Omega$. 
We note additionally that the decoherence effect can be mitigated through, e.g., bath engineering, to reduce the magnitude of this relative to the qubit coupling, i.e., $\gamma_f\ll J$~\cite{NaghilooMurch19nphys}.



There is an interesting additional robustness feature of the entanglement generation, namely that adding decoherence, e.g., in the form of dissipation on the $|f\rangle$ qubit levels, does not significantly change the time to reach maximal concurrence, although it does reduce the maximal value. 

Finally, we emphasize that our results demonstrate maximal concurrence at an optimal drive amplitude $\Omega^* > \Omega_{\rm EP}$ that ensures the system is close to but not located at the EP. A characteristic feature is that varying the drive away from this position, whether to larger or smaller frequencies, decreases the achievable entanglement. This implies that for a given value of $J$ with optimal $\Omega^*$, in the parameter region $\Omega_{\rm EP}<\Omega^*<\Omega$, there can be a competition between EP-enhanced decoherence~\cite{ChenMurch21prl} and EP-enhanced entanglement as the drive amplitude $\Omega$ is decreased toward the EP.